\documentclass[%
 reprint,
superscriptaddress,
 amsmath,amssymb,
 aps,
pre,
]{revtex4-2}
\usepackage{amsmath}%,amssymb} 
\usepackage{makeidx}
\usepackage{amsfonts}
\usepackage[ansinew]{inputenc}
\usepackage{subfigure}
\usepackage{hyperref}
\usepackage{cleveref}
\usepackage{graphicx}
\usepackage[outdir=./]{epstopdf}
\draft 

\begin{document}

\title{Enhanced proton acceleration from near-critical density targets employing intense lasers with mixed polarization}

\author{D. N. Gupta}
\author{S. Kumar}
\affiliation{Department of Physics and Astrophysics, University of Delhi, Delhi 110 007, India}
\author{S. Kar}
\affiliation{School of Mathematics and Physics, Queen's University Belfast, Belfast, BT7 1NN, UK}

\date{\today}

\begin{abstract}
 We demonstrate a scheme for enhanced proton acceleration from near-critical-density targets by splitting a laser pulse into a linearly and a circularly polarized laser pulse. The combination of two laser pulses generates a shock wave as well as hole boring effect at the front surface of the target. Protons at the front get combined acceleration from the two acceleration mechanisms. 2D-PIC simulation shows nearly two-fold enhancement in proton energy from this mechanism for a near-critical density target. The acceleration mechanism is also studied with different target density and thickness to ascertain the target parameters over which the acceleration process shows dominant behaviour. 
 
\end{abstract}


\maketitle 

\section{Introduction}
Collisionless shock wave formation is relevant in the field of astrophysical and space plasmas \cite{sagdeev1991,adriani2011}. The acceleration of cosmic particles through this shock wave is of fundamental importance \cite{spitkovsky2008,martins2009}. In laboratory, similar shock wave can be generated through laser plasma interaction. In recent year, the shock wave created by high-intensity laser is gaining wider interest as it can be an efficient mechanism for plasma based ion accelerators \cite{RevModPhys.85.751}. Energetic ion beams so generated have lots of applications in the field of laser fusion \cite{roth2001}, proton therapy \cite{linz2007}, isotope generation \cite{spencer2001}, radiography \cite{borghesi2008} etc. From laser solid interaction, proton beam can be generated through different mechanisms under different laser and plasma conditions \cite{fuchs2006,robson2007,PhysRevLett.109.185006,Nature2016,kumar2019}. At laser intensity as low as $1\times10^{18}~ W/cm^2 $, ions can be accelerated by the sheath field generated at the rear side of the target \cite{doi:10.1063/1.1333697}. This mechanism is known as the target normal sheath acceleration (TNSA) as ion beam is generated normal to the target with broader spectrum. Due to the limitation in laser intensity, most of the experimental work is based on TNSA mechanism \cite{PhysRevLett.85.2945,hegelich2005}. At higher laser intensity of $1\times10^{20-22}~ W/cm^2$, quasi-monoenergetic beams can be generated through the radiation pressure acceleration (RPA) mechanism \cite{Robinson_2008,klimo2008,kar2012}. This mechanism employs a circularly polarized laser pulse to push a double layer of electron and proton simultaneously inside the sub-micron thick target. Various numerical studies suggest that mono-energetic beam can be generated through RPA but it has limitations in terms of availability of ultra-intense laser pulse and handling of target with thickness in nanometre scale length. That's why there is a growing interest in collisionless shock wave acceleration \cite{silva2004,d2005proton} as it can be achieved with moderate laser intensity with target thickness in micron length.

Collisionless shock acceleration using particle simulations was first proposed by Denavit \emph{et. al.} \cite{PhysRevLett.69.3052}. When a high intensity laser pulse interacts with a solid target, it causes a density  spike to move inside the target. This density spike constitutes a electrostatic shock. The background ions in front of the shock get reflected due to the electric field associated with the shock and accelerated at twice the shock speed.  Basically, a shock wave is high amplitude ion acoustic wave. The laser pulse acts as piston and pushes the ions inside the target with some velocity. The velocity of the pushed ions can be determined using the momentum conservation relation \cite{silva2004,chen2007} i.e. $ (1+ \eta)I/c = m_in_iv_i^2 $ where $I$ is the laser intensity and $\eta$ represents the laser reflection efficiency. The ion velocity ($v_i)$ turns out to be $ v_i/c ={[(1+\eta)I/m_in_ic^3]}^{1/2} $. The shock wave velocity is close to the piston velocity i.e. $v_s \approx v_i$. In the frame associated with the shock velocity, the cold ions move with the shock velocity. If the maximum electrostatic potential ($\phi_{max})$ associated with the shock wave is greater than the the ions kinetic energy, than the particles get accelerated to twice the shock velocity \cite{he2007,stockem2013}. In general, for ions with charge $Z$, the condition is given by $Z_ie\phi_{max} \geq m_iv_s^2/2$. 

Various numerical \cite{silva2004,chen2007,liu2016,bhagawati2019} and experimental \cite{wei2004ion,zhang2015} works have been done to study shock wave acceleration mechanism. Silva \emph{et. al.} \cite{silva2004} have studied the shock wave acceleration under different laser and plasma parameters and gives certain threshold conditions. Sorasio \emph{et.al.} \cite{sorasio2006} have given the kinetic theory for shock formation from collision of two plasma slab having different properties. Chen \emph{et.al.} \cite{chen2007} have studied proton acceleration from shock wave under different laser and plasma conditions. In some of the studies, proton beam is generated with narrow energy spread using shock wave acceleration \cite{haberberger2012,fiuza2012}. Fiuza \emph{et al.} \cite{fiuza2013} have used tailored near critical density to generated proton beam with monoenergetic features. Both overdense target \cite{PhysRevLett.69.3052,silva2004,he2007} as well as near critical density (NCD) target \cite{fiuza2013,liu2016} are used to study shock wave acceleration. In NCD target, as laser can penetrate deeper in the target as well as there is efficient electron heating, thus it leads to a higher shock velocity.  With the advancement of high intensity infra-red laser, gas target can be used as near critical density target. The shock wave can be generated by linear as well as circularly polarized laser pulse. Palmer \emph{et. al.} \cite{palmer2011} have used circularly polarized $CO_2$ laser ($\lambda = 10 \mu m$) to produce radiation pressure shock in plasma and have shown nearly mono-energetic proton beams. Haberberger \emph{et.al.} \cite{haberberger2012} have used linear polarized $CO_2$ laser to generate thermal pressure shock and have demonstrated monoenergetic beam with NCD targets. Recently, Zhang \emph{et.al.} \cite{zhang2017} have used $0.8 \mu m$ wavelength circularly polarized laser pulse to drive a thermal pressure shock inside the target to accelerate protons. They have used oblique incident laser to create a density spike and generate energetic electrons to launch thermal shock. 

\begin{figure}
\centering
\includegraphics[scale = 0.3]{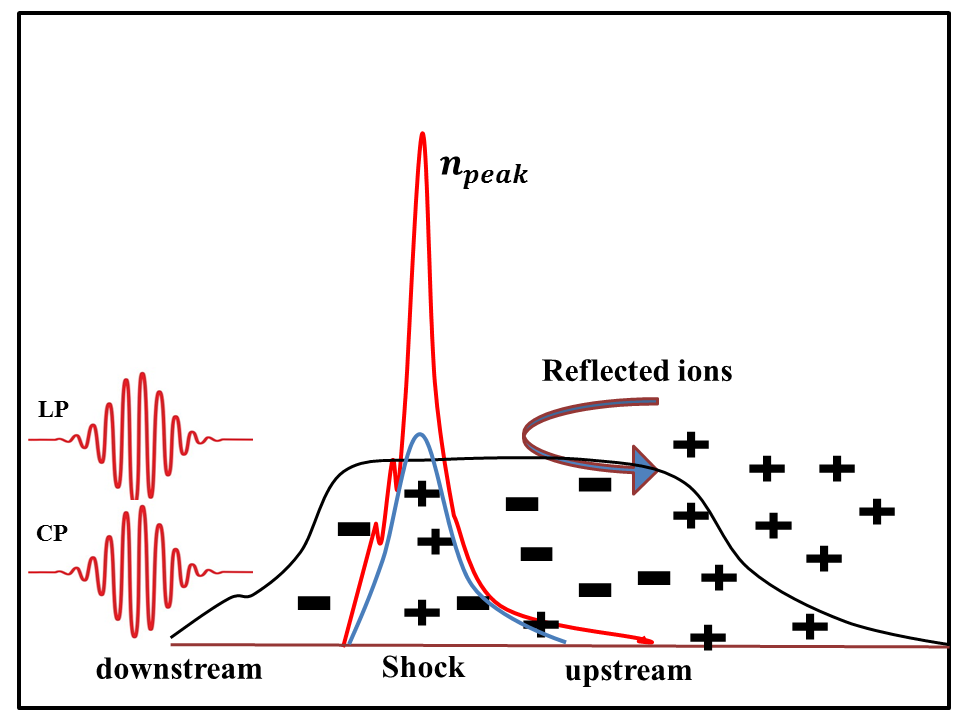}
\caption{Schematic of proton acceleration mechanism by mixed laser polarizations. The red and blue peaks represent the proton density peak arises due to the interaction of circular and linear polarized laser pulses, respectively. The reflected ions represent the protons accelerated through the combined effect of two mechanisms.}
\label{I}
\end{figure}

An effective shock can be generated if the laser interaction at the front of the target causes a strong piston like push as well as significant electron heating. The ion acoustic wave speed depends on the electron temperature given by the relation $c_s = \sqrt{k_BT_e/m_i}$. Thus in order to generate a high amplitude ion acoustic wave, we need effective electron heating at the front surface of the target. Generally, linearly polarized laser pulse is used to generate thermal shock wave and circularly polarized laser pulse is used to produce radiation pressure shock at the front surface of the target. In this paper, we propose to investigate the proton acceleration with a combination of linearly and circularly polarized laser pulse using a NCD target. With the application of mixed polarization, both the initial push and electron heating can be substantially achieved to take shock wave inside the plasma. Along with the shock wave formation, the circularly polarized laser pulse can causes hole boring effect at the front surface of the target. The schematic of proposed proton acceleration mechanism with mixed polarization is shown in Fig. \ref {I}. The red density peak presents the hole boring effect and the blue density peak corresponds to the shock wave. With the help of 2D-PIC simulation, we will show that the proton acceleration as combination of these two effects leads to a higher proton energy. We also study that the interaction of NCD with linear and circular polarized laser pulse individually so that a
comparison can be possible with the double laser pulse case and a complete picture can be presented. It has to be noted that while comparing, the total energy of the two laser pulse in mixed polarization is taken same as that of single linear and circular polarized laser pulse.

\section{Simulation Details}
To carry out PIC simulation, we use fully relativistic PIC code "EPOCH" \cite{arber2015}. We consider a simulation window of size $ 50\times 40~\mu m$ in x-y plane. The laser pulse originates from the left side of the simulation box and the peak intensity of a single laser pulse (linear polarized) is taken to be $ 1.36 \times 10^{20}~W/cm^2 $, which corresponds to the laser intensity parameter $ a_0 = 10 $. The wavelength of the laser pulse is considered to be $\lambda = 1~\mu m$. The spot radius and pulse duration of the laser pulse is taken to be $6~\mu m$ and $ 40~fs $, respectively. The energy of the laser pulse corresponding to these laser parameters is turns out to be $6~J$. In order to use the combination of linear and circularly polarized laser pulse, we divide the laser energy between two polarizations. Thus the comparison between single and double laser is done with the same total laser energy in both cases. The target is placed at the center of the simulation box with thickness $3~\mu m $ and density $10 ~n_c$, where $n_c$ represents the critical density given by $n_c = 1.1 \times 10^{21}/\lambda_{\mu m}^2~cm^{-3}$. The grid size for the simulation is taken to be $10~nm$ and $40~nm$ in $x$ and $y$ directions, respectively. The plasma target consists of electrons and protons as separate species. All the boundaries of the simulation box are set to be absorbing. All the simulations are run upto a sufficient time such that we reach a saturation point.

\section{Simulation Results and Discussion}
\begin{figure*}
\begin{center}
\includegraphics[scale = 0.23]{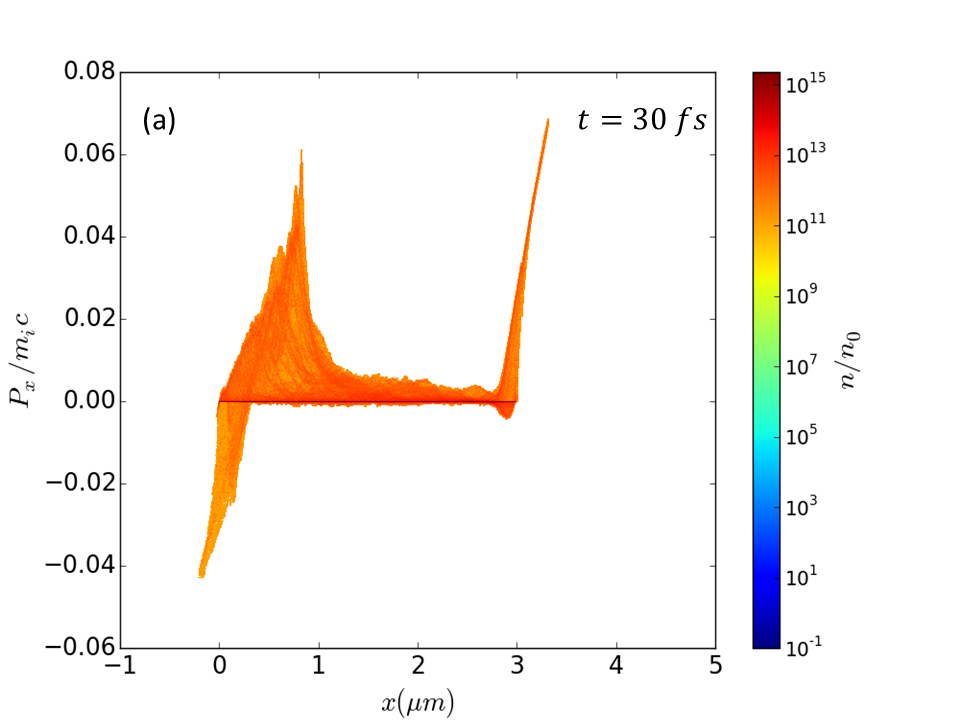}
\includegraphics[scale = 0.23]{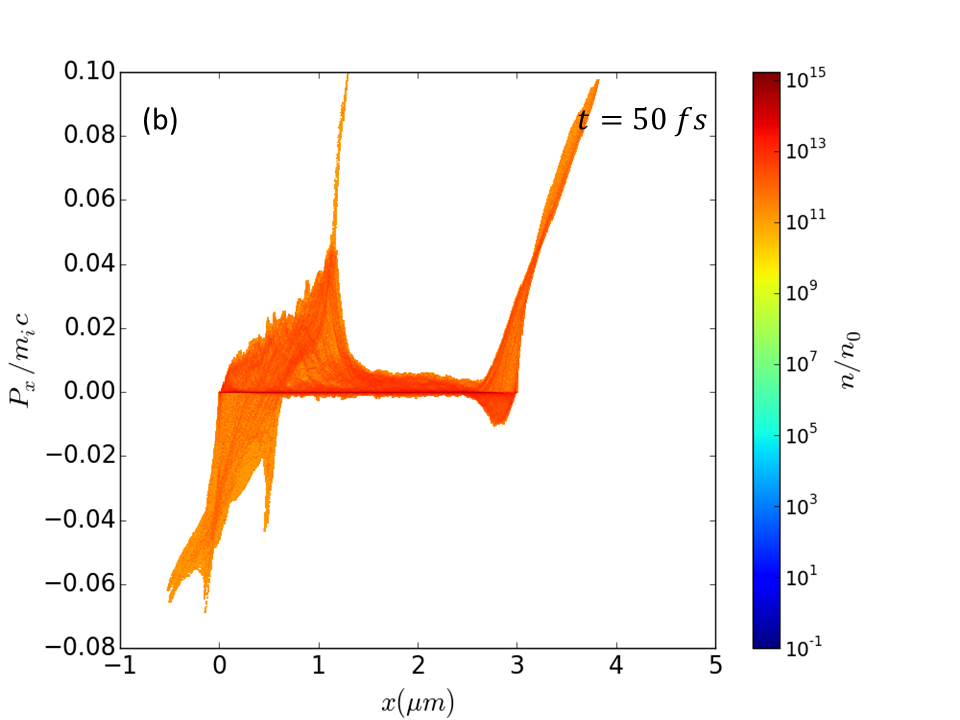}
\includegraphics[scale = 0.22]{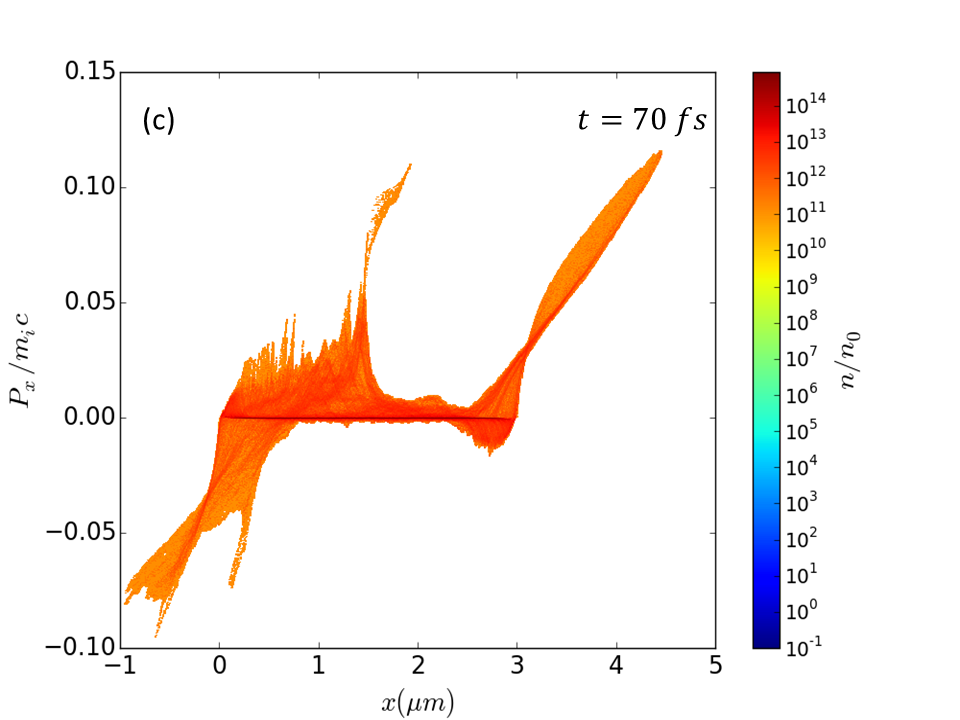}
\includegraphics[scale = 0.23]{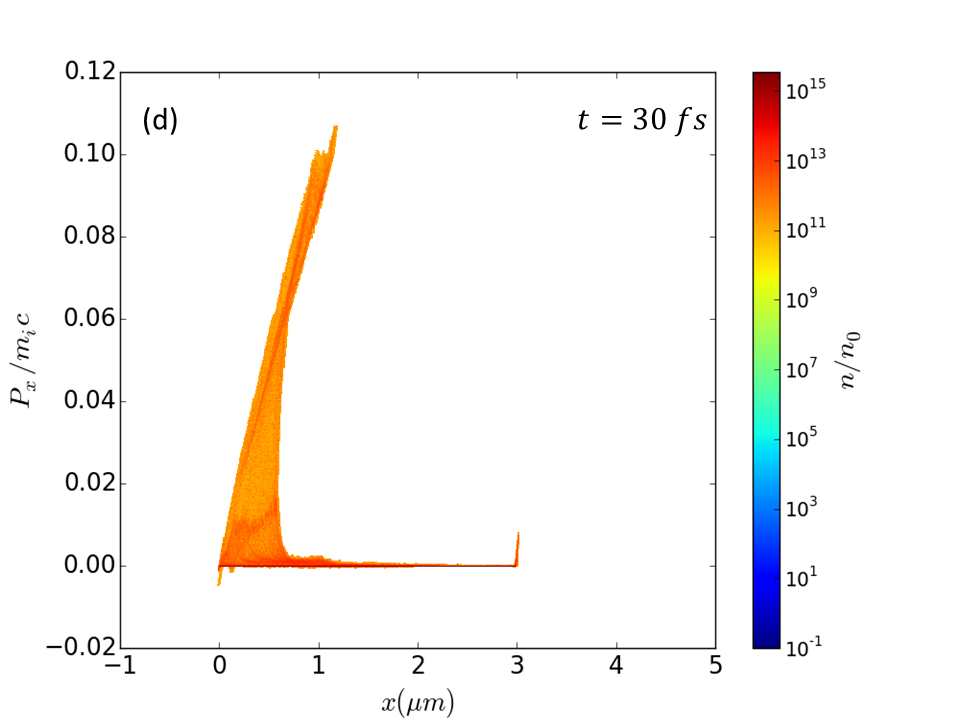}
\includegraphics[scale = 0.23]{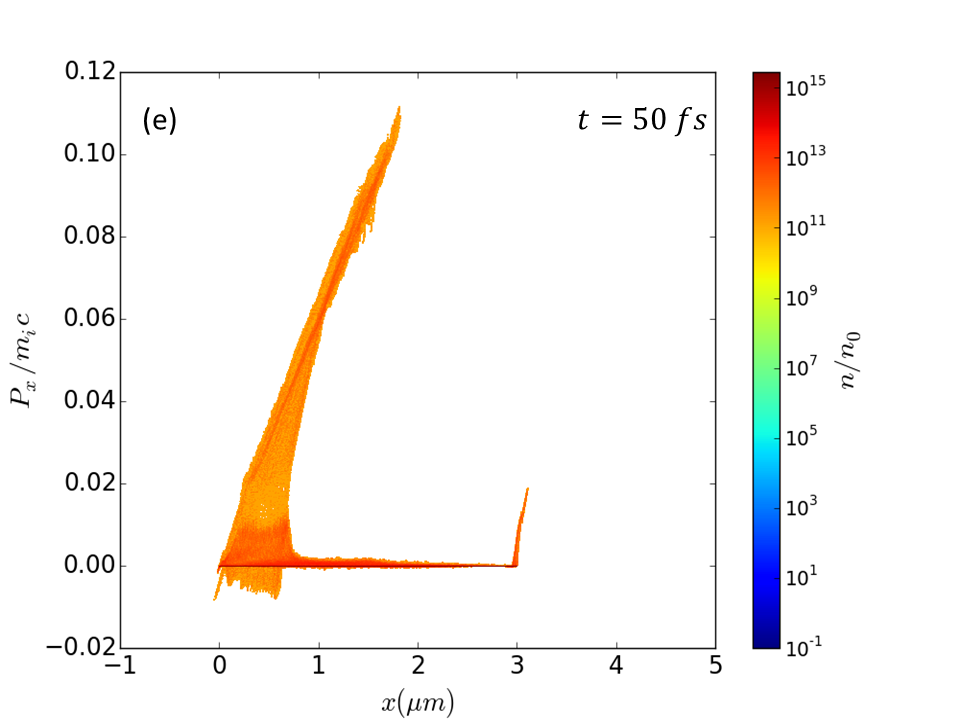}
\includegraphics[scale = 0.23]{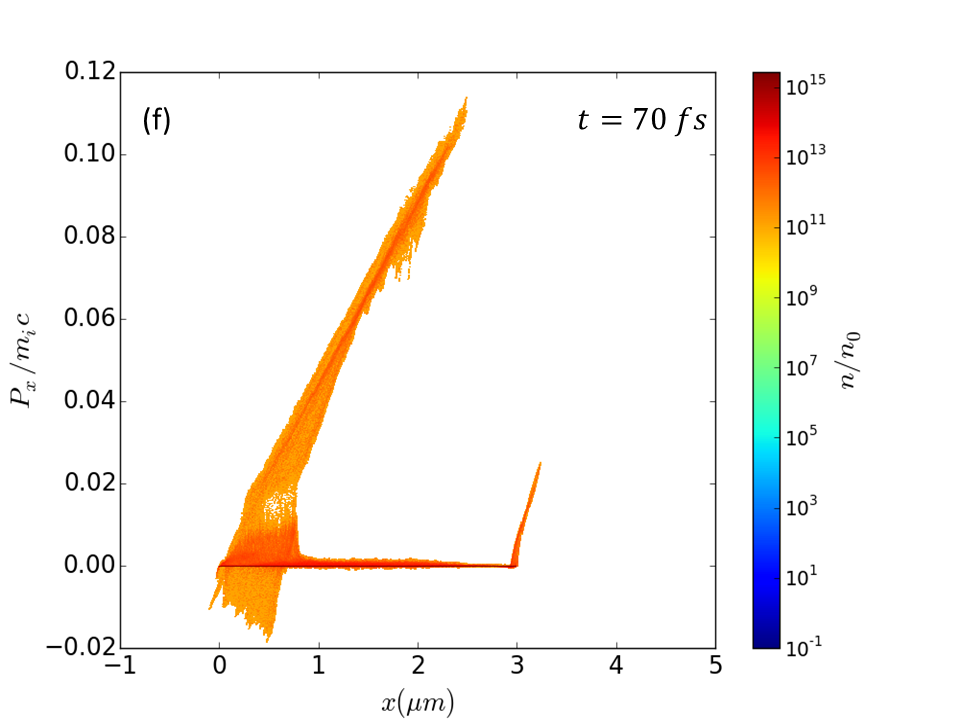}
\includegraphics[scale = 0.23]{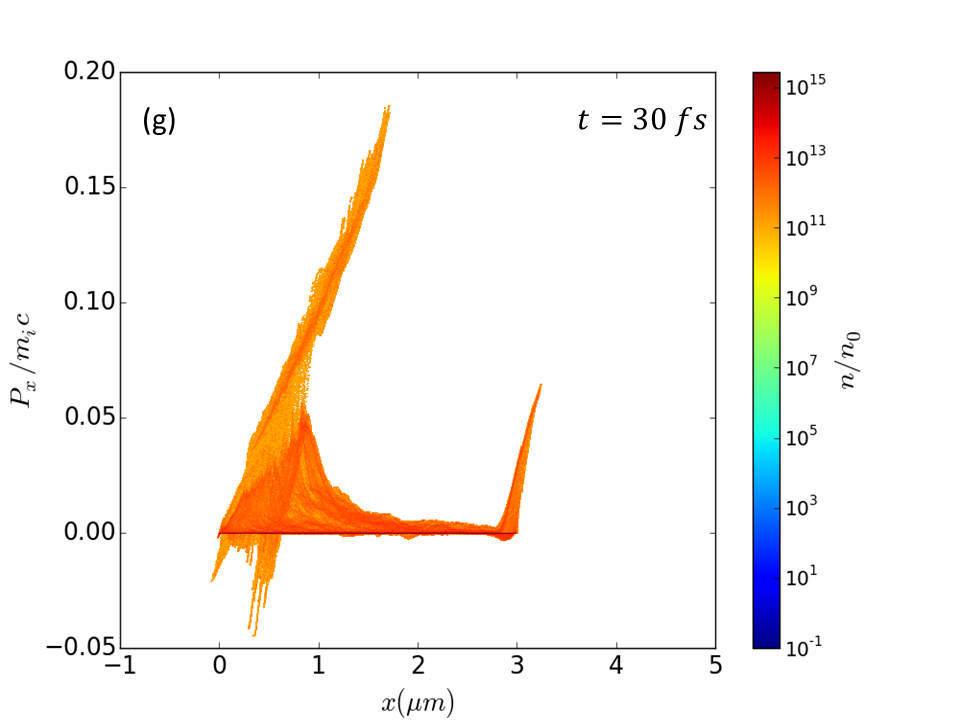}
\includegraphics[scale = 0.23]{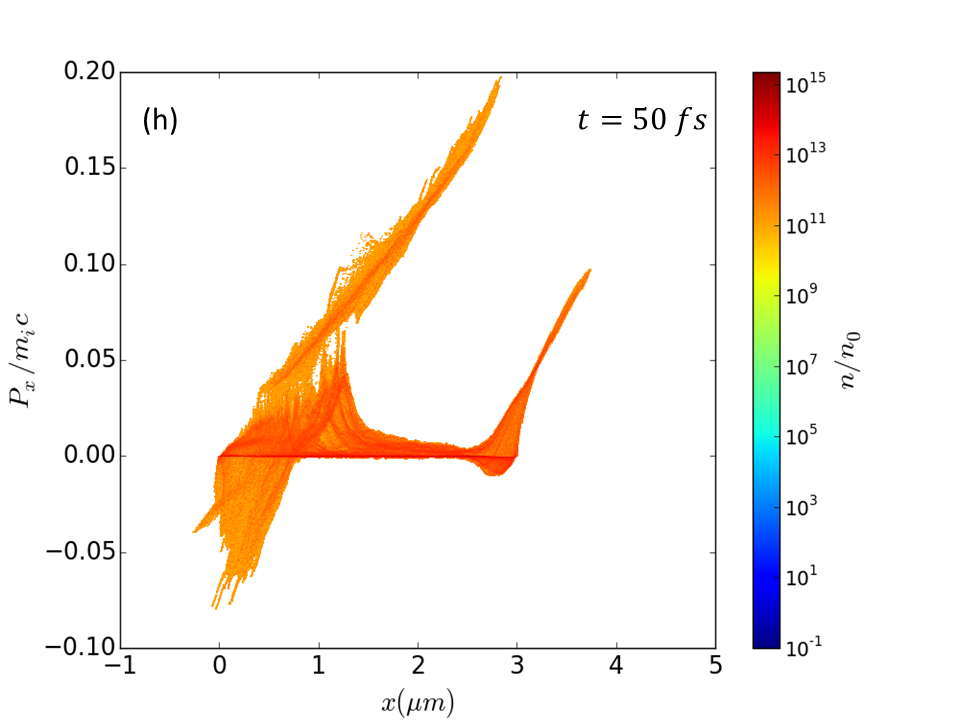}
\includegraphics[scale = 0.22]{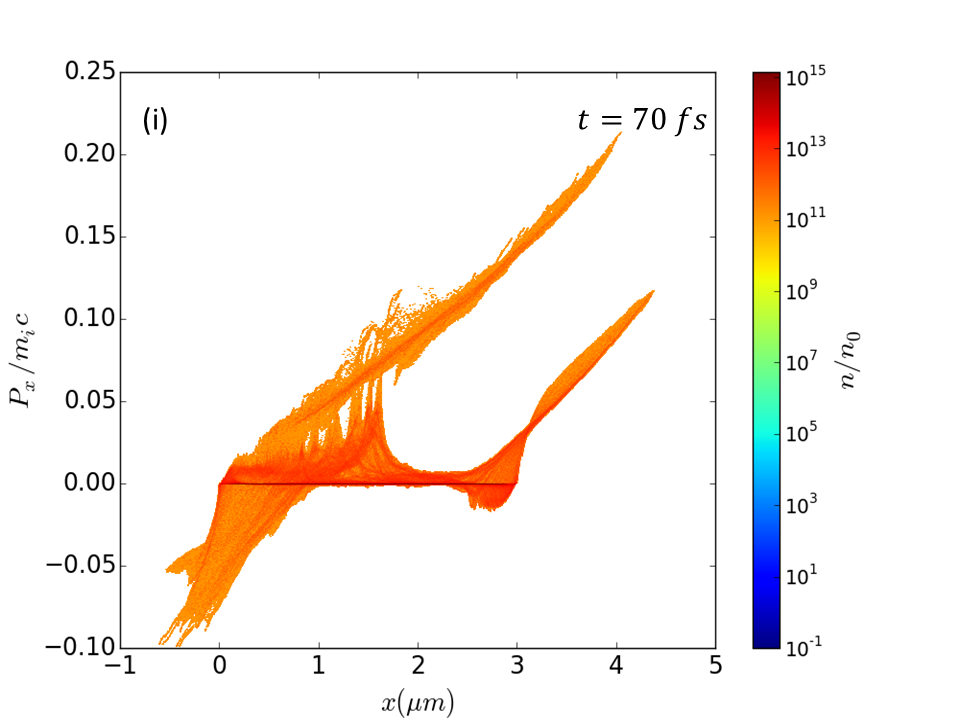}
\caption{Phase-space plots for protons at different times after the laser peak interacts with the front surface of the target for (a-c) linearly polarized, (d-f) circularly polarized, and (g-i) linearly$+$ circularly polarized laser pulse. The target thickness and density are $3~\mu m$ and $10n_c$, respectively for each case. Also, the total laser energy on target in each case in $6~J$ with spot radius $6~\mu m$ and pulse duration $40~fs.$}
\label{1}
\end{center}
\end{figure*}

Using the simulation parameters discussed in previous section, initially, we run the simulations with linear polarized laser pulse. The laser intensity and energy is taken to be $a_0 = 10 $ and $E = 6 ~J $, respectively. The laser pulse normally incident on the target of thickness $3~\mu m $ and density $n= 10n_c $. The ultra-intense laser causes hole boring action as well as electron heating at the front surface of the target. Thus, a shock wave is created and it moves inside the target. The heated electrons move to the rear surface of the target and form the sheath field. The protons get accelerated due to the shock wave as well as due to the sheath field. In Fig. \ref{1}    (a-c), we show the phase-space for protons along the laser propagation direction at three different time at $t = 30~fs, 50~fs$  and $ 70~fs $ after the laser peak interacts with the target. The normalized momentum of the shock wave turns out to be around $0.05$ and it moves inside the target with almost constant velocity. There is evidence of proton acceleration from shock wave as represented by phase space plot at $t = 70~fs.$ However, as we focus on the rear side of the target at the same time, the protons get accelerated due to the sheath field are more energetic as compared to the shock wave. The protons at the rear surface get further accelerated at latter time and indeed get much higher energy as compared to shock accelerated protons. In this case, the TNSA mechanism dominates over the shock acceleration mechanism. As the shock velocity is low, the shock accelerated protons reach the rear side of the target at latter time, so protons do not get any significant energy gain from the sheath field. There is effectively no coupling between the shock acceleration and the TNSA mechanism. 
In the next simulation, we studied the effect of circularly polarized laser pulse on a NCD target having thickness in a few micron scale. The energy of the circularly polarized laser is considered to be same as that of the linearly polarized laser pulse i.e. $6~J$. The corresponding phase-space plots for protons for this case are shown in Figs. \ref{1} (d-f). As circularly polarized laser pulse inhibits the formation of hot electrons, it shows a minimal proton acceleration at the rear surface of the target. At the front surface of the target, the process of the radiation pressure acceleration dominates. The hole boring effect takes place at the front surface and the moving density surface accelerates the protons at the front surface of the target.  As we compare Figs. 2(c) and 2(f), energy of the protons accelerated from the shock is almost the same. Due to the dominant TNSA effect in the linearly polarized case, proton energy comes out to be higher at later time.
\begin{figure}
\begin{center}
\includegraphics[scale = 0.35]{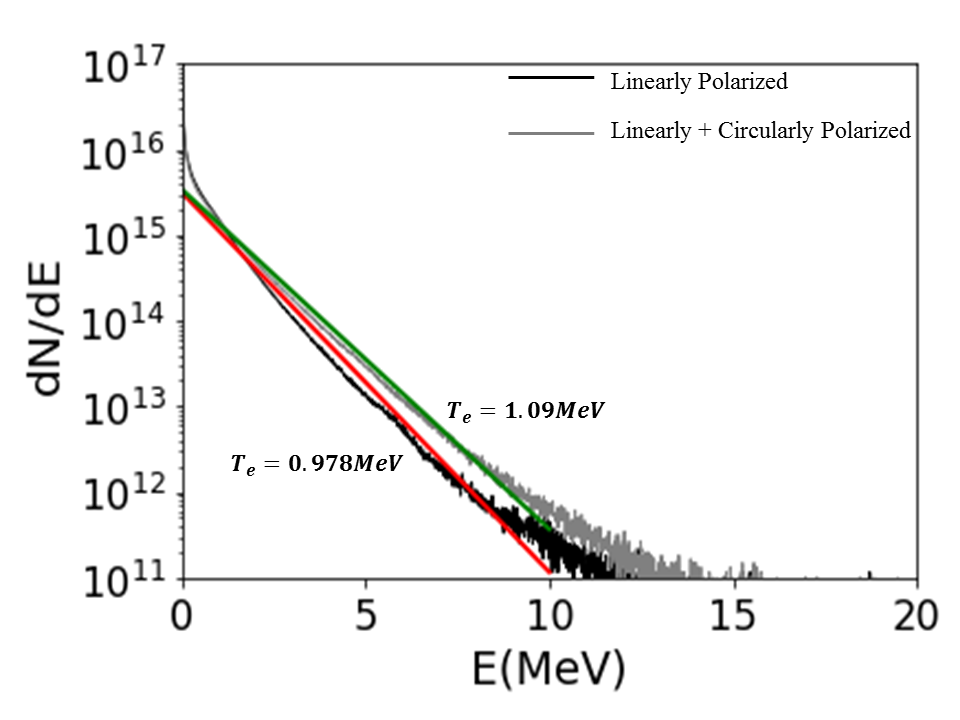}
\caption{Electron energy distribution at $70~fs $ time after the laser pulse peak interacts with the front surface of the target. The target thickness and density are $3~\mu m$ and $10n_c$, respectively. The laser spot radius and duration are $6~\mu m$ and $40~fs$, respectively. }
\label{2}
\end{center}
\end{figure}

In the next simulations, we divided the laser energy of $6~J$ between two laser pulses, one with linear polarization and other with circular polarization. Both the linearly and circularly polarized laser pulses are incident on the target surface simultaneously. As the oscillating component of the ponderomotive force vanishes in the case of circularly polarized laser pulse, it efficiently pushes the target surface inside. The linearly polarized laser pulse, as usual, generates energetic electrons, which produces a strong TNSA field at the target rear surface. At the front surface, the combined effect of hole boring and shock acceleration takes place. Fig. \ref{1}(g-i) shows the phase-space plots for this case. The phase-space plots show the formation of shock wave as well as the radiation pressure effect at the front surface of the target. This combined effect leads to the enhanced acceleration of protons. Protons accelerated by the hole boring effect are further accelerated after getting reflected from the shock front. The normalized shock momentum is around $0.05$ with application of double laser pulse. The shock wave velocity is almost same as compared to the single linearly polarized laser pulse. Due to the higher proton energy at the front surface, it reaches to the target rear surface at earlier time. As the linearly polarized laser pulse is also used along with circularly polarized pulse, thus the sheath field is also formed at the rear side of the target. Therefore, the shock accelerated protons are further accelerated by the electrostatic sheath field as they reach to the rear side of the target. The results show that the two acceleration mechanisms can be coupled effectively using linearly and circularly polarized laser pulse simultaneously. We have also measured electron temperature for linear and mixed polarization case. We have determined the electron temperature from its energy distribution at time $70~fs$ after the laser peak interacts with the target. The electron temperature for the two cases are nearly same and it is about $0.978~MeV $ and $1.02~MeV$ for linear polarization and double laser pulse cases, respectively. The temperature is almost same because due to the radiation pressure, the target surface get deformed and causes sufficient electron heating in double pulse case. 
\begin{figure}
\begin{center}
\includegraphics[scale = 0.37]{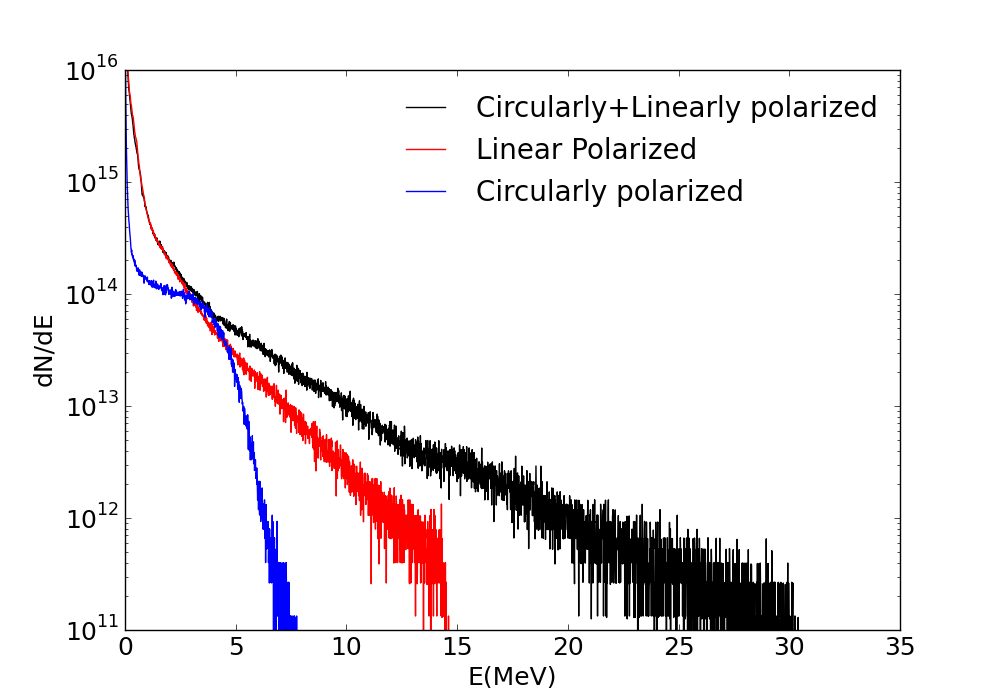}
\caption{Proton energy spectrum for three different cases, i.e. linearly polarized laser pulse, circularly polarized laser pulse, linearly+circularly polarized laser pulse. The target thickness used for these simulations is $3~\mu m$ with density $10n_c$. The laser spot radius and pulse duration are $6~\mu m$ and $40~fs$, respectively.}
\label{2}
\end{center}
\end{figure}

\begin{figure}
\begin{center}
\includegraphics[scale = 0.37]{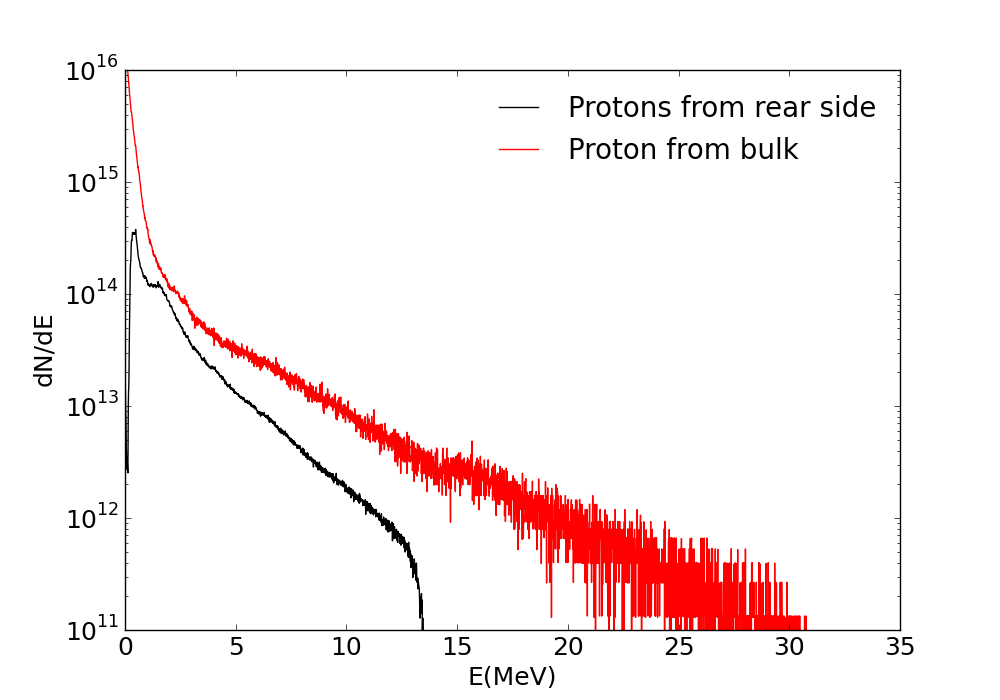}
\caption{Energy spectrum for bulk and rear surface protons using the case of linear$+$ circularly polarized laser pulse with laser spot radius of $6~\mu m$ and pulse duration of $40~fs$ for both  the laser pulses. The bulk target thickness and contamination layer thickness are $ 3~\mu m $ and $100~nm$, respectively with density $10n_c$ of each.}
\label{3}
\end{center}
\end{figure}

The shock enhanced proton acceleration can be confirmed by looking at the proton energy spectra as shown in Fig. \ref{2}. The energy spectra is observed at $370~fs $ after the laser peak interacts with the front surface of the target as around this time proton energy reaches its saturation value. The maximum proton cut off energy reaches upto $ 30~MeV $ for shock dominated process. For independent cases of linearly and circularly polarized laser pulses, we are getting maximum proton energy of $15~MeV $ and $8~MeV$, respectively. In all these simulations, the laser energy is same, thus the maximum transfer of the laser energy to protons is achieved by splitting the single laser into linear and circularly polarized laser pulses. The proton cut off energy is almost doubled in this case as compared to the former case. In order to get further understanding about the energy acquired by the bulk and rear surface protons, we also plotted the energy spectrum for bulk and rear surface protons separately. In order to differentiate between protons in the bulk and at the rear surface of the target, we added an extra proton layer of $ 100~nm $ at the rear surface and tracked them separately. Protons in this layer can act as contamination layer generally present in lab targets. The corresponding energy spectrum is shown in Fig. \ref{3}. The bulk protons, which are initially accelerated by the combined effect of hole boring and shock wave are more energetic. The mechanism of TNSA is also evident at the rear surface of the target and the maximum proton energy comes out to be around $15~MeV$. 

\begin{figure}
\begin{center}
\includegraphics[scale =0.35]{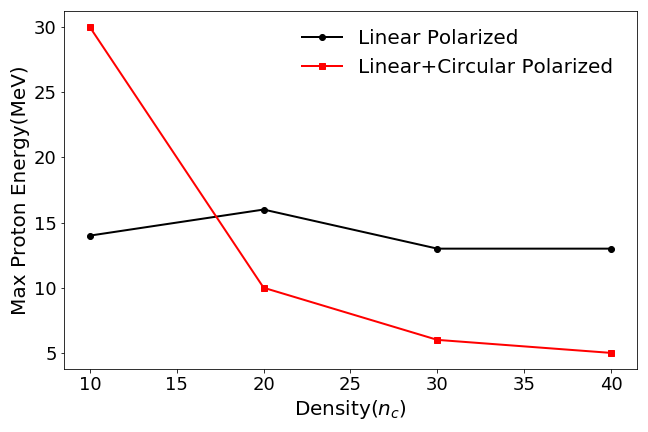}
\caption{Maximum Proton energy as a function of target density for linearly and linearly + circularly polarized laser pulse with target thickness fixed at $3~\mu m$.}
\label{4}
\end{center}
\end{figure}

The process of shock wave acceleration as well as TNSA mechanism depends on the target density. We run another set of simulations with varying target density to study the dominant acceleration mechanism at different densities. Other laser and target parameters were kept unchanged as discussed previously. The simulation results in Fig. \ref{4} show that, with combination of linear and circularly polarized laser pulses, the proton acceleration at the front surface is the dominant mechanism at low density. As the density increases to $20n_c$, there is a steep reduction in peak energy of the accelerated protons. This is primarily because the hole boring velocity decreases as it depends inversely on the density. At higher density of $30n_c$ and $40n_c$, TNSA as expected is the dominant acceleration mechanism. If we consider the case of single linearly polarized laser pulse, the proton energy is maximum at all density except at near critical density of $10n_c$. Shock acceleration coupled with the RPA mechanism is only dominant around the critical density to generate energetic protons. Thus, to generate efficient hole boring and shock wave as well as to get energetic proton beam, near critical density (NCD) targets are more favourable. As we employ laser pulse of same energy for all the cases, hence, the laser absorption is higher for near critical density and it is efficiently transferred to the protons. 

\begin{figure}
\begin{center}
\includegraphics[scale =0.35]{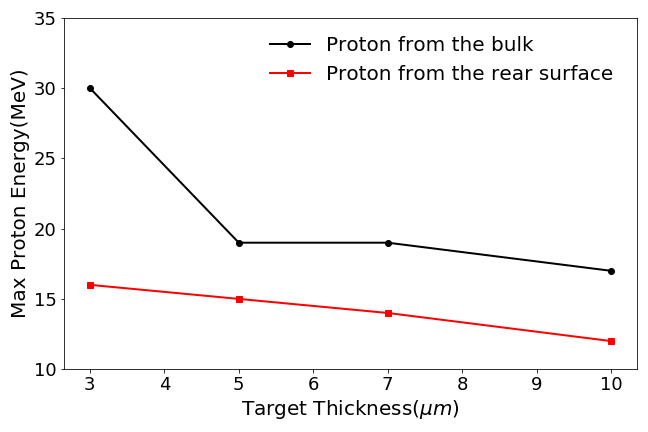}
\caption{Maximum Proton energy as a function of target thickness for linearly + circularly polarized laser pulse with spot radius $5~\mu m$ and pulse duration $40~fs$ for both the laser pulses. The contamination layer thickness is fixed at $100~nm$ and bulk target thickness is varied. The density is $10n_c$ for each layer.}
\label{5}
\end{center}
\end{figure}

As the density variation study shows that the effect of double laser pulse is dominant at near critical density, we fix the density of the target at $10nc$ and study the effect of target thickness on proton cut-off energy. We consider the case of double laser pulse (i.e. combination of linearly and circularly polarized laser pulses) and varied the target thickness between $3~\mu m$ and $10~\mu m$. Again, in order to see the effect of proton acceleration from the rear surface, we consider a proton layer of $ 100~nm $ at the rear surface and track it independently. Fig. \ref{5} shows the variation of cut-off energy of protons from the bulk and protons from the rear surface as the target thickness increases. With increasing the target thickness, proton cut-off energy from front surface gets decreased. It is mainly because the accelerated protons take longer to reach to the rear side of the target, the sheath field has already died out and cannot further accelerate the protons. As we focused on rear side of the target, the variation of target thickness in this thickness range has no significant effect on proton cut-off energy. This shows that the interaction with double laser pulse has higher laser energy absorption. As a result, higher energy is only possible with thin NCD target with thickness in sub-micron range.

\begin{figure}
\begin{center}
\includegraphics[scale =0.35]{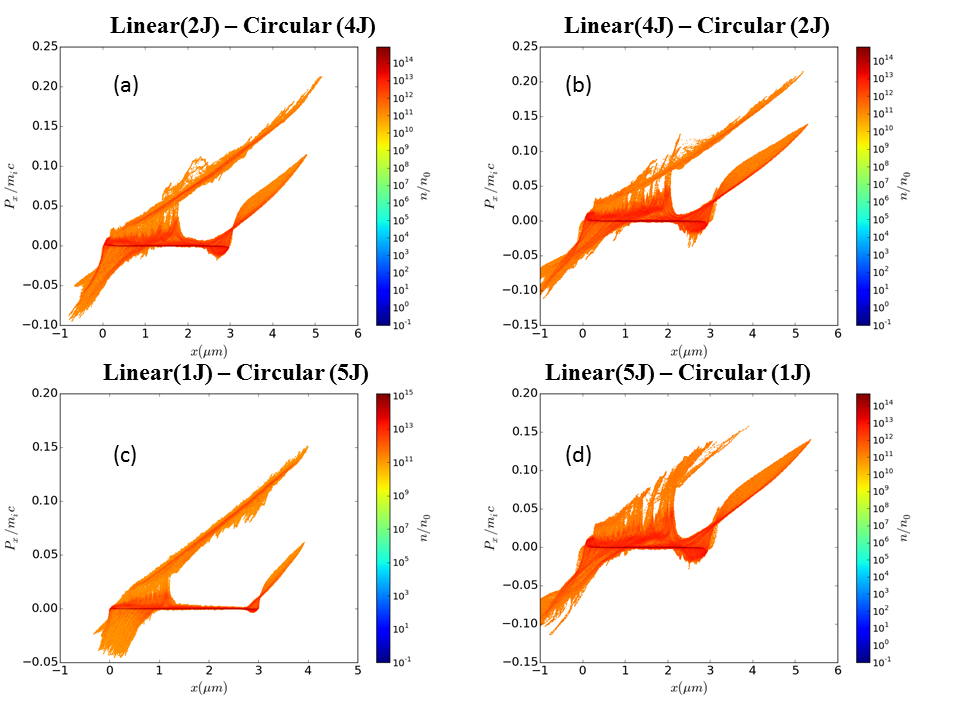}
\caption{Proton phase plot with different distributions of energy between linear and circularly polarized laser pulses. In each case, the target thickness is $3~\mu m$ with density $10n_c$.}
\label{6}
\end{center}
\end{figure}

For all the above simulation involving double laser pulse, we have consider equal energy \emph{i.e} $3$ J in both laser pulses. In order to determine the effect of different distributions of laser energy in linear and circularly polarized laser pulses, we run simulations with different combinations of laser energy. We have taken energy in the combinations of $(2J-4J), ~(4J-2J), ~(1J-5J),~ (5J-1J)$. where the first term represents the energy in linear polarized laser pulse and second term represents the laser energy in circularly polarized laser pulse. Fig. \ref{6} shows the proton phase-space plot for different distributions of energy between two laser pulses at $t=90~fs$ (after the laser peak interacts with the target). The phase-space plot suggests that the hole-boring effect dominates in all the cases except the case ($5J-1J$). In Fig. \ref{6}(d) the thermal shock wave is a dominant acceleration mechanism at the front surface of the target. Also, the shock wave velocity is higher in this case as compared to all other combination of energy distribution. The phase space plot also suggests that, for the maximum proton energy, energy distribution should be such that a strong hole boring effect takes place along with a significant thermal shock wave. Therefore, using mixed polarization, higher proton energy can be achieved with unequal distribution of energy between two laser pulses (with linear polarized laser having lesser energy as compared to the circularly polarized laser pulse).

Finally, we have studied the effect of increasing the energy of the laser pulse. When a single laser pulse with circular polarization with energy of $3~J$ interacts with a plasma target of thickness $3~\mu m $ and density $10n_c$, the maximum proton energy is around $4~MeV$. As we change the polarization of the laser pulse to linear while keeping the laser energy constant, the proton cut-off energy is around $10~MeV$. This shows that with these target parameters, linear polarization gives higher proton energy and the effect of TNSA is dominant than the RPA mechanism. Now in order to get higher proton energy, it is reasonable to increase the laser energy. As we double the laser energy to $6~J$ for the case of linear polarization, proton cut off energy increases to $15~MeV$. There is increase of only $3~MeV$ when the laser energy is doubled. On contrary, we get almost two fold increase in proton energy by using $3~J $ of linear and $3~J $ of circular polarized pulse. Thus application of double laser pulse with mixed polarization can be an effective scheme for proton acceleration.

\begin{figure}
\begin{center}
\includegraphics[scale = 0.37]{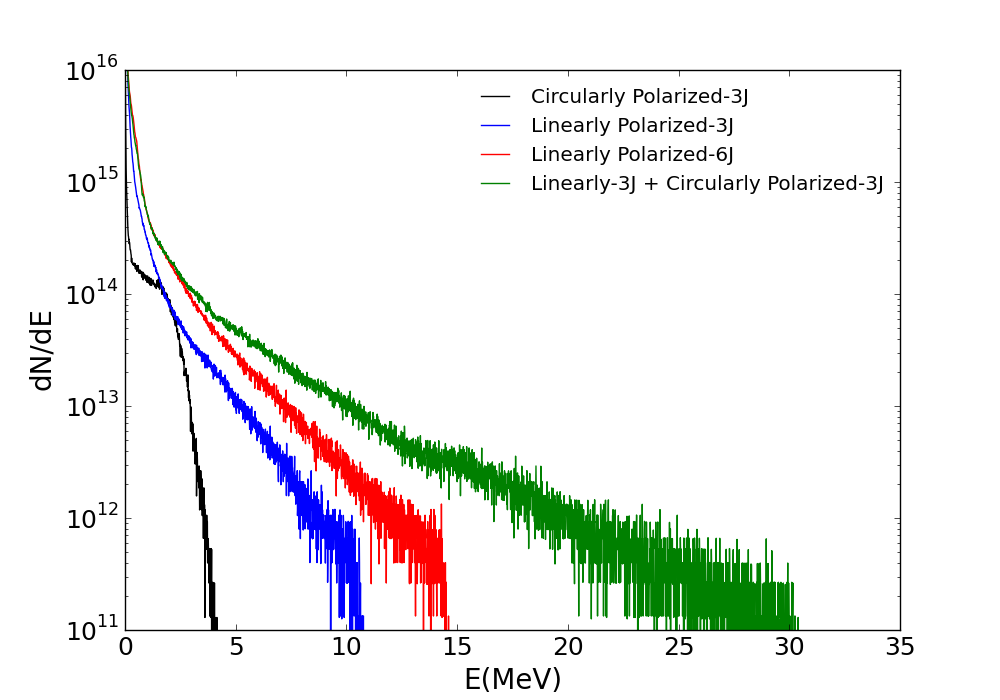}
\caption{Proton energy distribution with combination of laser pulse and respective laser pulse energy.}
\label{7}
\end{center}
\end{figure}
\section{conclusion}
The results presented here show a new scheme of proton acceleration by enhancing the front surface proton acceleration using linearly and circularly polarized laser pulses together. The shock enhanced proton acceleration is investigated from the near critical density target with sub-micron thickness. The idea behind this investigation was to split a single laser pulse into two laser pulses with linearly and circularly polarization.
The use of single laser pulse generates proton beam with energy around $15~MeV $. The dominant acceleration mechanism is target normal sheath acceleration with a small effect of shock acceleration at the front of the target in that case. On the other hand, we found that, if the single laser pulse is split into a linear and circularly polarized laser pulse (keeping the energy conservation hold), the proton energy gets almost doubled fold, which is around $ 30~MeV.$ The observation of the phase-space plots suggested that the dominant acceleration mechanism is hole boring and is booted by the shock wave acceleration. The proton are dominantly accelerated at the front side. There is possibility of further acceleration at the rear side of the target by the longitudinal sheath field, if the proton from the front reaches there just in time to experience the non-vanishing sheath field. This study have also reported that the shock wave acceleration is effective with the low-density target. As the target density is increased, the TNSA mechanism dominates. Also, the target thickness variation studies have shown that as the target thickness increases, the proton energy have decreased. Thus, at a low-density target of thickness in $1\sim 3 \mu m$ range with a constant supply of laser energy, this study suggest to split the laser pulse in two pulses of different polarizations in order to produce more energetic proton beams. 

\section{Acknowledgements}
The authors would like to thank Department of Physics and Astrophysics, University of Delhi, Delhi, India for providing high performance cluster machine to carry out the simulation work. The author Saurabh Kumar would also like to thank the University Grant Commission (Government of India) for providing Senior Research Fellowship for the Ph.D. program.

\section*{Disclosures}
The author declares no conflicts of interest. The data that support the findings of this study are available from the corresponding author upon reasonable request.


\begin{thebibliography}{37}%
\makeatletter
\providecommand \@ifxundefined [1]{%
 \@ifx{#1\undefined}
}%
\providecommand \@ifnum [1]{%
 \ifnum #1\expandafter \@firstoftwo
 \else \expandafter \@secondoftwo
 \fi
}%
\providecommand \@ifx [1]{%
 \ifx #1\expandafter \@firstoftwo
 \else \expandafter \@secondoftwo
 \fi
}%
\providecommand \natexlab [1]{#1}%
\providecommand \enquote  [1]{``#1''}%
\providecommand \bibnamefont  [1]{#1}%
\providecommand \bibfnamefont [1]{#1}%
\providecommand \citenamefont [1]{#1}%
\providecommand \href@noop [0]{\@secondoftwo}%
\providecommand \href [0]{\begingroup \@sanitize@url \@href}%
\providecommand \@href[1]{\@@startlink{#1}\@@href}%
\providecommand \@@href[1]{\endgroup#1\@@endlink}%
\providecommand \@sanitize@url [0]{\catcode `\\12\catcode `\$12\catcode
  `\&12\catcode `\#12\catcode `\^12\catcode `\_12\catcode `\%12\relax}%
\providecommand \@@startlink[1]{}%
\providecommand \@@endlink[0]{}%
\providecommand \url  [0]{\begingroup\@sanitize@url \@url }%
\providecommand \@url [1]{\endgroup\@href {#1}{\urlprefix }}%
\providecommand \urlprefix  [0]{URL }%
\providecommand \Eprint [0]{\href }%
\providecommand \doibase [0]{https://doi.org/}%
\providecommand \selectlanguage [0]{\@gobble}%
\providecommand \bibinfo  [0]{\@secondoftwo}%
\providecommand \bibfield  [0]{\@secondoftwo}%
\providecommand \translation [1]{[#1]}%
\providecommand \BibitemOpen [0]{}%
\providecommand \bibitemStop [0]{}%
\providecommand \bibitemNoStop [0]{.\EOS\space}%
\providecommand \EOS [0]{\spacefactor3000\relax}%
\providecommand \BibitemShut  [1]{\csname bibitem#1\endcsname}%
\let\auto@bib@innerbib\@empty
%</preamble>
\bibitem [{\citenamefont {Sagdeev}\ and\ \citenamefont
  {Kennel}(1991)}]{sagdeev1991}%
  \BibitemOpen
  \bibfield  {author} {\bibinfo {author} {\bibfnamefont {R.~Z.}\ \bibnamefont
  {Sagdeev}}\ and\ \bibinfo {author} {\bibfnamefont {C.~F.}\ \bibnamefont
  {Kennel}},\ }\href@noop {} {\bibfield  {journal} {\bibinfo  {journal}
  {Scientific American}\ }\textbf {\bibinfo {volume} {264}},\ \bibinfo {pages}
  {106} (\bibinfo {year} {1991})}\BibitemShut {NoStop}%
\bibitem [{\citenamefont {Adriani}\ \emph {et~al.}(2011)\citenamefont
  {Adriani}, \citenamefont {Barbarino}, \citenamefont {Bazilevskaya},
  \citenamefont {Bellotti}, \citenamefont {Boezio}, \citenamefont {Bogomolov},
  \citenamefont {Bonechi}, \citenamefont {Bongi}, \citenamefont {Bonvicini},
  \citenamefont {Borisov}, \citenamefont {Bottai}, \citenamefont {Bruno},
  \citenamefont {Cafagna}, \citenamefont {Campana}, \citenamefont {Carbone},
  \citenamefont {Carlson}, \citenamefont {Casolino}, \citenamefont
  {Castellini}, \citenamefont {Consiglio}, \citenamefont {Pascale},
  \citenamefont {Santis}, \citenamefont {Simone}, \citenamefont {Felice},
  \citenamefont {Galper}, \citenamefont {Gillard}, \citenamefont
  {Grishantseva}, \citenamefont {Jerse}, \citenamefont {Karelin}, \citenamefont
  {Koldashov}, \citenamefont {Krutkov}, \citenamefont {Kvashnin}, \citenamefont
  {Leonov}, \citenamefont {Malakhov}, \citenamefont {Malvezzi}, \citenamefont
  {Marcelli}, \citenamefont {Mayorov}, \citenamefont {Menn}, \citenamefont
  {Mikhailov}, \citenamefont {Mocchiutti}, \citenamefont {Monaco},
  \citenamefont {Mori}, \citenamefont {Nikonov}, \citenamefont {Osteria},
  \citenamefont {Palma}, \citenamefont {Papini}, \citenamefont {Pearce},
  \citenamefont {Picozza}, \citenamefont {Pizzolotto}, \citenamefont {Ricci},
  \citenamefont {Ricciarini}, \citenamefont {Rossetto}, \citenamefont {Sarkar},
  \citenamefont {Simon}, \citenamefont {Sparvoli}, \citenamefont {Spillantini},
  \citenamefont {Stozhkov}, \citenamefont {Vacchi}, \citenamefont {Vannuccini},
  \citenamefont {Vasilyev}, \citenamefont {Voronov}, \citenamefont {Yurkin},
  \citenamefont {Wu}, \citenamefont {Zampa}, \citenamefont {Zampa},\ and\
  \citenamefont {Zverev}}]{adriani2011}%
  \BibitemOpen
  \bibfield  {author} {\bibinfo {author} {\bibfnamefont {O.}~\bibnamefont
  {Adriani}}, \bibinfo {author} {\bibfnamefont {G.~C.}\ \bibnamefont
  {Barbarino}}, \bibinfo {author} {\bibfnamefont {G.~A.}\ \bibnamefont
  {Bazilevskaya}}, \bibinfo {author} {\bibfnamefont {R.}~\bibnamefont
  {Bellotti}}, \bibinfo {author} {\bibfnamefont {M.}~\bibnamefont {Boezio}},
  \bibinfo {author} {\bibfnamefont {E.~A.}\ \bibnamefont {Bogomolov}}, \bibinfo
  {author} {\bibfnamefont {L.}~\bibnamefont {Bonechi}}, \bibinfo {author}
  {\bibfnamefont {M.}~\bibnamefont {Bongi}}, \bibinfo {author} {\bibfnamefont
  {V.}~\bibnamefont {Bonvicini}}, \bibinfo {author} {\bibfnamefont
  {S.}~\bibnamefont {Borisov}}, \bibinfo {author} {\bibfnamefont
  {S.}~\bibnamefont {Bottai}}, \bibinfo {author} {\bibfnamefont
  {A.}~\bibnamefont {Bruno}}, \bibinfo {author} {\bibfnamefont
  {F.}~\bibnamefont {Cafagna}}, \bibinfo {author} {\bibfnamefont
  {D.}~\bibnamefont {Campana}}, \bibinfo {author} {\bibfnamefont
  {R.}~\bibnamefont {Carbone}}, \bibinfo {author} {\bibfnamefont
  {P.}~\bibnamefont {Carlson}}, \bibinfo {author} {\bibfnamefont
  {M.}~\bibnamefont {Casolino}}, \bibinfo {author} {\bibfnamefont
  {G.}~\bibnamefont {Castellini}}, \bibinfo {author} {\bibfnamefont
  {L.}~\bibnamefont {Consiglio}}, \bibinfo {author} {\bibfnamefont {M.~P.~D.}\
  \bibnamefont {Pascale}}, \bibinfo {author} {\bibfnamefont {C.~D.}\
  \bibnamefont {Santis}}, \bibinfo {author} {\bibfnamefont {N.~D.}\
  \bibnamefont {Simone}}, \bibinfo {author} {\bibfnamefont {V.~D.}\
  \bibnamefont {Felice}}, \bibinfo {author} {\bibfnamefont {A.~M.}\
  \bibnamefont {Galper}}, \bibinfo {author} {\bibfnamefont {W.}~\bibnamefont
  {Gillard}}, \bibinfo {author} {\bibfnamefont {L.}~\bibnamefont
  {Grishantseva}}, \bibinfo {author} {\bibfnamefont {G.}~\bibnamefont {Jerse}},
  \bibinfo {author} {\bibfnamefont {A.~V.}\ \bibnamefont {Karelin}}, \bibinfo
  {author} {\bibfnamefont {S.~V.}\ \bibnamefont {Koldashov}}, \bibinfo {author}
  {\bibfnamefont {S.~Y.}\ \bibnamefont {Krutkov}}, \bibinfo {author}
  {\bibfnamefont {A.~N.}\ \bibnamefont {Kvashnin}}, \bibinfo {author}
  {\bibfnamefont {A.}~\bibnamefont {Leonov}}, \bibinfo {author} {\bibfnamefont
  {V.}~\bibnamefont {Malakhov}}, \bibinfo {author} {\bibfnamefont
  {V.}~\bibnamefont {Malvezzi}}, \bibinfo {author} {\bibfnamefont
  {L.}~\bibnamefont {Marcelli}}, \bibinfo {author} {\bibfnamefont {A.~G.}\
  \bibnamefont {Mayorov}}, \bibinfo {author} {\bibfnamefont {W.}~\bibnamefont
  {Menn}}, \bibinfo {author} {\bibfnamefont {V.~V.}\ \bibnamefont {Mikhailov}},
  \bibinfo {author} {\bibfnamefont {E.}~\bibnamefont {Mocchiutti}}, \bibinfo
  {author} {\bibfnamefont {A.}~\bibnamefont {Monaco}}, \bibinfo {author}
  {\bibfnamefont {N.}~\bibnamefont {Mori}}, \bibinfo {author} {\bibfnamefont
  {N.}~\bibnamefont {Nikonov}}, \bibinfo {author} {\bibfnamefont
  {G.}~\bibnamefont {Osteria}}, \bibinfo {author} {\bibfnamefont
  {F.}~\bibnamefont {Palma}}, \bibinfo {author} {\bibfnamefont
  {P.}~\bibnamefont {Papini}}, \bibinfo {author} {\bibfnamefont
  {M.}~\bibnamefont {Pearce}}, \bibinfo {author} {\bibfnamefont
  {P.}~\bibnamefont {Picozza}}, \bibinfo {author} {\bibfnamefont
  {C.}~\bibnamefont {Pizzolotto}}, \bibinfo {author} {\bibfnamefont
  {M.}~\bibnamefont {Ricci}}, \bibinfo {author} {\bibfnamefont {S.~B.}\
  \bibnamefont {Ricciarini}}, \bibinfo {author} {\bibfnamefont
  {L.}~\bibnamefont {Rossetto}}, \bibinfo {author} {\bibfnamefont
  {R.}~\bibnamefont {Sarkar}}, \bibinfo {author} {\bibfnamefont
  {M.}~\bibnamefont {Simon}}, \bibinfo {author} {\bibfnamefont
  {R.}~\bibnamefont {Sparvoli}}, \bibinfo {author} {\bibfnamefont
  {P.}~\bibnamefont {Spillantini}}, \bibinfo {author} {\bibfnamefont {Y.~I.}\
  \bibnamefont {Stozhkov}}, \bibinfo {author} {\bibfnamefont {A.}~\bibnamefont
  {Vacchi}}, \bibinfo {author} {\bibfnamefont {E.}~\bibnamefont {Vannuccini}},
  \bibinfo {author} {\bibfnamefont {G.}~\bibnamefont {Vasilyev}}, \bibinfo
  {author} {\bibfnamefont {S.~A.}\ \bibnamefont {Voronov}}, \bibinfo {author}
  {\bibfnamefont {Y.~T.}\ \bibnamefont {Yurkin}}, \bibinfo {author}
  {\bibfnamefont {J.}~\bibnamefont {Wu}}, \bibinfo {author} {\bibfnamefont
  {G.}~\bibnamefont {Zampa}}, \bibinfo {author} {\bibfnamefont
  {N.}~\bibnamefont {Zampa}},\ and\ \bibinfo {author} {\bibfnamefont {V.~G.}\
  \bibnamefont {Zverev}},\ }\href@noop {} {\bibfield  {journal} {\bibinfo
  {journal} {Science}\ }\textbf {\bibinfo {volume} {332}},\ \bibinfo {pages}
  {69} (\bibinfo {year} {2011})}\BibitemShut {NoStop}%
\bibitem [{\citenamefont {Spitkovsky}(2008)}]{spitkovsky2008}%
  \BibitemOpen
  \bibfield  {author} {\bibinfo {author} {\bibfnamefont {A.}~\bibnamefont
  {Spitkovsky}},\ }\href@noop {} {\bibfield  {journal} {\bibinfo  {journal}
  {The Astrophysical Journal Letters}\ }\textbf {\bibinfo {volume} {682}},\
  \bibinfo {pages} {L5} (\bibinfo {year} {2008})}\BibitemShut {NoStop}%
\bibitem [{\citenamefont {Martins}\ \emph {et~al.}(2009)\citenamefont
  {Martins}, \citenamefont {Fonseca}, \citenamefont {Silva},\ and\
  \citenamefont {Mori}}]{martins2009}%
  \BibitemOpen
  \bibfield  {author} {\bibinfo {author} {\bibfnamefont {S.}~\bibnamefont
  {Martins}}, \bibinfo {author} {\bibfnamefont {R.}~\bibnamefont {Fonseca}},
  \bibinfo {author} {\bibfnamefont {L.}~\bibnamefont {Silva}},\ and\ \bibinfo
  {author} {\bibfnamefont {W.}~\bibnamefont {Mori}},\ }\href@noop {} {\bibfield
   {journal} {\bibinfo  {journal} {The Astrophysical Journal Letters}\ }\textbf
  {\bibinfo {volume} {695}},\ \bibinfo {pages} {L189} (\bibinfo {year}
  {2009})}\BibitemShut {NoStop}%
\bibitem [{\citenamefont {Macchi}\ \emph {et~al.}(2013)\citenamefont {Macchi},
  \citenamefont {Borghesi},\ and\ \citenamefont {Passoni}}]{RevModPhys.85.751}%
  \BibitemOpen
  \bibfield  {author} {\bibinfo {author} {\bibfnamefont {A.}~\bibnamefont
  {Macchi}}, \bibinfo {author} {\bibfnamefont {M.}~\bibnamefont {Borghesi}},\
  and\ \bibinfo {author} {\bibfnamefont {M.}~\bibnamefont {Passoni}},\
  }\href@noop {} {\bibfield  {journal} {\bibinfo  {journal} {Review of Modern
  Physics}\ }\textbf {\bibinfo {volume} {85}},\ \bibinfo {pages} {751}
  (\bibinfo {year} {2013})}\BibitemShut {NoStop}%
\bibitem [{\citenamefont {Roth}\ \emph {et~al.}(2001)\citenamefont {Roth},
  \citenamefont {Cowan}, \citenamefont {Key}, \citenamefont {Hatchett},
  \citenamefont {Brown}, \citenamefont {Fountain}, \citenamefont {Johnson},
  \citenamefont {Pennington}, \citenamefont {Snavely}, \citenamefont {Wilks},
  \citenamefont {Yasuike}, \citenamefont {Ruhl}, \citenamefont {Pegoraro},
  \citenamefont {Bulanov}, \citenamefont {Campbell}, \citenamefont {Perry},\
  and\ \citenamefont {Powell}}]{roth2001}%
  \BibitemOpen
  \bibfield  {author} {\bibinfo {author} {\bibfnamefont {M.}~\bibnamefont
  {Roth}}, \bibinfo {author} {\bibfnamefont {T.~E.}\ \bibnamefont {Cowan}},
  \bibinfo {author} {\bibfnamefont {M.~H.}\ \bibnamefont {Key}}, \bibinfo
  {author} {\bibfnamefont {S.~P.}\ \bibnamefont {Hatchett}}, \bibinfo {author}
  {\bibfnamefont {C.}~\bibnamefont {Brown}}, \bibinfo {author} {\bibfnamefont
  {W.}~\bibnamefont {Fountain}}, \bibinfo {author} {\bibfnamefont
  {J.}~\bibnamefont {Johnson}}, \bibinfo {author} {\bibfnamefont {D.~M.}\
  \bibnamefont {Pennington}}, \bibinfo {author} {\bibfnamefont {R.~A.}\
  \bibnamefont {Snavely}}, \bibinfo {author} {\bibfnamefont {S.~C.}\
  \bibnamefont {Wilks}}, \bibinfo {author} {\bibfnamefont {K.}~\bibnamefont
  {Yasuike}}, \bibinfo {author} {\bibfnamefont {H.}~\bibnamefont {Ruhl}},
  \bibinfo {author} {\bibfnamefont {F.}~\bibnamefont {Pegoraro}}, \bibinfo
  {author} {\bibfnamefont {S.~V.}\ \bibnamefont {Bulanov}}, \bibinfo {author}
  {\bibfnamefont {E.~M.}\ \bibnamefont {Campbell}}, \bibinfo {author}
  {\bibfnamefont {M.~D.}\ \bibnamefont {Perry}},\ and\ \bibinfo {author}
  {\bibfnamefont {H.}~\bibnamefont {Powell}},\ }\href@noop {} {\bibfield
  {journal} {\bibinfo  {journal} {Phys. Rev. Lett.}\ }\textbf {\bibinfo
  {volume} {86}},\ \bibinfo {pages} {436} (\bibinfo {year} {2001})}\BibitemShut
  {NoStop}%
\bibitem [{\citenamefont {Linz}\ and\ \citenamefont {Alonso}(2007)}]{linz2007}%
  \BibitemOpen
  \bibfield  {author} {\bibinfo {author} {\bibfnamefont {U.}~\bibnamefont
  {Linz}}\ and\ \bibinfo {author} {\bibfnamefont {J.}~\bibnamefont {Alonso}},\
  }\href@noop {} {\bibfield  {journal} {\bibinfo  {journal} {Physical Review
  Special Topics-Accelerators and Beams}\ }\textbf {\bibinfo {volume} {10}},\
  \bibinfo {pages} {094801} (\bibinfo {year} {2007})}\BibitemShut {NoStop}%
\bibitem [{\citenamefont {Spencer}\ \emph {et~al.}(2001)\citenamefont
  {Spencer}, \citenamefont {Ledingham}, \citenamefont {Singhal}, \citenamefont
  {McCanny}, \citenamefont {McKenna}, \citenamefont {Clark}, \citenamefont
  {Krushelnick}, \citenamefont {Zepf}, \citenamefont {Beg}, \citenamefont
  {Tatarakis}, \citenamefont {Dangor}, \citenamefont {Norreys}, \citenamefont
  {Clarke}, \citenamefont {Allott},\ and\ \citenamefont {Ross}}]{spencer2001}%
  \BibitemOpen
  \bibfield  {author} {\bibinfo {author} {\bibfnamefont {I.}~\bibnamefont
  {Spencer}}, \bibinfo {author} {\bibfnamefont {K.}~\bibnamefont {Ledingham}},
  \bibinfo {author} {\bibfnamefont {R.}~\bibnamefont {Singhal}}, \bibinfo
  {author} {\bibfnamefont {T.}~\bibnamefont {McCanny}}, \bibinfo {author}
  {\bibfnamefont {P.}~\bibnamefont {McKenna}}, \bibinfo {author} {\bibfnamefont
  {E.}~\bibnamefont {Clark}}, \bibinfo {author} {\bibfnamefont
  {K.}~\bibnamefont {Krushelnick}}, \bibinfo {author} {\bibfnamefont
  {M.}~\bibnamefont {Zepf}}, \bibinfo {author} {\bibfnamefont {F.}~\bibnamefont
  {Beg}}, \bibinfo {author} {\bibfnamefont {M.}~\bibnamefont {Tatarakis}},
  \bibinfo {author} {\bibfnamefont {A.}~\bibnamefont {Dangor}}, \bibinfo
  {author} {\bibfnamefont {P.}~\bibnamefont {Norreys}}, \bibinfo {author}
  {\bibfnamefont {R.}~\bibnamefont {Clarke}}, \bibinfo {author} {\bibfnamefont
  {R.}~\bibnamefont {Allott}},\ and\ \bibinfo {author} {\bibfnamefont
  {I.}~\bibnamefont {Ross}},\ }\href@noop {} {\bibfield  {journal} {\bibinfo
  {journal} {Nuclear Instruments and Methods in Physics Research Section B:
  Beam Interactions with Materials and Atoms}\ }\textbf {\bibinfo {volume}
  {183}},\ \bibinfo {pages} {449} (\bibinfo {year} {2001})}\BibitemShut
  {NoStop}%
\bibitem [{\citenamefont {Borghesi}\ \emph {et~al.}(2008)\citenamefont
  {Borghesi}, \citenamefont {Bigongiari}, \citenamefont {Kar}, \citenamefont
  {Macchi}, \citenamefont {Romagnani}, \citenamefont {Audebert}, \citenamefont
  {Fuchs}, \citenamefont {Toncian}, \citenamefont {Willi}, \citenamefont
  {Bulanov}, \citenamefont {Mackinnon},\ and\ \citenamefont
  {C}}]{borghesi2008}%
  \BibitemOpen
  \bibfield  {author} {\bibinfo {author} {\bibfnamefont {M.}~\bibnamefont
  {Borghesi}}, \bibinfo {author} {\bibfnamefont {A.}~\bibnamefont
  {Bigongiari}}, \bibinfo {author} {\bibfnamefont {S.}~\bibnamefont {Kar}},
  \bibinfo {author} {\bibfnamefont {A.}~\bibnamefont {Macchi}}, \bibinfo
  {author} {\bibfnamefont {L.}~\bibnamefont {Romagnani}}, \bibinfo {author}
  {\bibfnamefont {P.}~\bibnamefont {Audebert}}, \bibinfo {author}
  {\bibfnamefont {J.}~\bibnamefont {Fuchs}}, \bibinfo {author} {\bibfnamefont
  {T.}~\bibnamefont {Toncian}}, \bibinfo {author} {\bibfnamefont
  {O.}~\bibnamefont {Willi}}, \bibinfo {author} {\bibfnamefont {S.~V.}\
  \bibnamefont {Bulanov}}, \bibinfo {author} {\bibfnamefont {A.~J.}\
  \bibnamefont {Mackinnon}},\ and\ \bibinfo {author} {\bibfnamefont {G.~J.}\
  \bibnamefont {C}},\ }\href@noop {} {\bibfield  {journal} {\bibinfo  {journal}
  {Plasma Physics and Controlled Fusion}\ }\textbf {\bibinfo {volume} {50}},\
  \bibinfo {pages} {124040} (\bibinfo {year} {2008})}\BibitemShut {NoStop}%
\bibitem [{\citenamefont {Fuchs}\ \emph {et~al.}(2006)\citenamefont {Fuchs},
  \citenamefont {Antici}, \citenamefont {Humires}, \citenamefont {Lefebvre},
  \citenamefont {Borghesi}, \citenamefont {Brambrink}, \citenamefont
  {Cecchetti}, \citenamefont {Kaluza}, \citenamefont {Malka}, \citenamefont
  {Manclossi},\ and\ \citenamefont {Meyroneinc}}]{fuchs2006}%
  \BibitemOpen
  \bibfield  {author} {\bibinfo {author} {\bibfnamefont {J.}~\bibnamefont
  {Fuchs}}, \bibinfo {author} {\bibfnamefont {P.}~\bibnamefont {Antici}},
  \bibinfo {author} {\bibfnamefont {E.}~\bibnamefont {Humires}}, \bibinfo
  {author} {\bibfnamefont {E.}~\bibnamefont {Lefebvre}}, \bibinfo {author}
  {\bibfnamefont {M.}~\bibnamefont {Borghesi}}, \bibinfo {author}
  {\bibfnamefont {E.}~\bibnamefont {Brambrink}}, \bibinfo {author}
  {\bibfnamefont {C.}~\bibnamefont {Cecchetti}}, \bibinfo {author}
  {\bibfnamefont {M.}~\bibnamefont {Kaluza}}, \bibinfo {author} {\bibfnamefont
  {V.}~\bibnamefont {Malka}}, \bibinfo {author} {\bibfnamefont
  {M.}~\bibnamefont {Manclossi}},\ and\ \bibinfo {author} {\bibfnamefont
  {S.}~\bibnamefont {Meyroneinc}},\ }\href@noop {} {\bibfield  {journal}
  {\bibinfo  {journal} {Nature Physics}\ }\textbf {\bibinfo {volume} {2}},\
  \bibinfo {pages} {48} (\bibinfo {year} {2006})}\BibitemShut {NoStop}%
\bibitem [{\citenamefont {Robson}\ \emph {et~al.}(2007)\citenamefont {Robson},
  \citenamefont {Simpson}, \citenamefont {Clarke}, \citenamefont {Ledingham},
  \citenamefont {Lindau}, \citenamefont {Lundh}, \citenamefont {McCanny},
  \citenamefont {Mora}, \citenamefont {Neely}, \citenamefont {Wahlstr{\"o}m},\
  and\ \citenamefont {Zepf}}]{robson2007}%
  \BibitemOpen
  \bibfield  {author} {\bibinfo {author} {\bibfnamefont {L.}~\bibnamefont
  {Robson}}, \bibinfo {author} {\bibfnamefont {P.}~\bibnamefont {Simpson}},
  \bibinfo {author} {\bibfnamefont {R.~J.}\ \bibnamefont {Clarke}}, \bibinfo
  {author} {\bibfnamefont {K.~W.}\ \bibnamefont {Ledingham}}, \bibinfo {author}
  {\bibfnamefont {F.}~\bibnamefont {Lindau}}, \bibinfo {author} {\bibfnamefont
  {O.}~\bibnamefont {Lundh}}, \bibinfo {author} {\bibfnamefont
  {T.}~\bibnamefont {McCanny}}, \bibinfo {author} {\bibfnamefont
  {P.}~\bibnamefont {Mora}}, \bibinfo {author} {\bibfnamefont {D.}~\bibnamefont
  {Neely}}, \bibinfo {author} {\bibfnamefont {C.-G.}\ \bibnamefont
  {Wahlstr{\"o}m}},\ and\ \bibinfo {author} {\bibfnamefont {M.}~\bibnamefont
  {Zepf}},\ }\href@noop {} {\bibfield  {journal} {\bibinfo  {journal} {Nature
  Physics}\ }\textbf {\bibinfo {volume} {3}},\ \bibinfo {pages} {58} (\bibinfo
  {year} {2007})}\BibitemShut {NoStop}%
\bibitem [{\citenamefont {Kar}\ \emph {et~al.}(2012)\citenamefont {Kar},
  \citenamefont {Kakolee}, \citenamefont {Qiao}, \citenamefont {Macchi},
  \citenamefont {Cerchez}, \citenamefont {Doria}, \citenamefont {Geissler},
  \citenamefont {McKenna}, \citenamefont {Neely}, \citenamefont {Osterholz},
  \citenamefont {Prasad}, \citenamefont {Quinn}, \citenamefont {Ramakrishna},
  \citenamefont {Sarri}, \citenamefont {Willi}, \citenamefont {Yuan},
  \citenamefont {Zepf},\ and\ \citenamefont
  {Borghesi}}]{PhysRevLett.109.185006}%
  \BibitemOpen
  \bibfield  {author} {\bibinfo {author} {\bibfnamefont {S.}~\bibnamefont
  {Kar}}, \bibinfo {author} {\bibfnamefont {K.~F.}\ \bibnamefont {Kakolee}},
  \bibinfo {author} {\bibfnamefont {B.}~\bibnamefont {Qiao}}, \bibinfo {author}
  {\bibfnamefont {A.}~\bibnamefont {Macchi}}, \bibinfo {author} {\bibfnamefont
  {M.}~\bibnamefont {Cerchez}}, \bibinfo {author} {\bibfnamefont
  {D.}~\bibnamefont {Doria}}, \bibinfo {author} {\bibfnamefont
  {M.}~\bibnamefont {Geissler}}, \bibinfo {author} {\bibfnamefont
  {P.}~\bibnamefont {McKenna}}, \bibinfo {author} {\bibfnamefont
  {D.}~\bibnamefont {Neely}}, \bibinfo {author} {\bibfnamefont
  {J.}~\bibnamefont {Osterholz}}, \bibinfo {author} {\bibfnamefont
  {R.}~\bibnamefont {Prasad}}, \bibinfo {author} {\bibfnamefont
  {K.}~\bibnamefont {Quinn}}, \bibinfo {author} {\bibfnamefont
  {B.}~\bibnamefont {Ramakrishna}}, \bibinfo {author} {\bibfnamefont
  {G.}~\bibnamefont {Sarri}}, \bibinfo {author} {\bibfnamefont
  {O.}~\bibnamefont {Willi}}, \bibinfo {author} {\bibfnamefont {X.~Y.}\
  \bibnamefont {Yuan}}, \bibinfo {author} {\bibfnamefont {M.}~\bibnamefont
  {Zepf}},\ and\ \bibinfo {author} {\bibfnamefont {M.}~\bibnamefont
  {Borghesi}},\ }\href {https://doi.org/10.1103/PhysRevLett.109.185006}
  {\bibfield  {journal} {\bibinfo  {journal} {Physical Review Letters}\
  }\textbf {\bibinfo {volume} {109}},\ \bibinfo {pages} {185006} (\bibinfo
  {year} {2012})}\BibitemShut {NoStop}%
\bibitem [{\citenamefont {Kar}\ \emph {et~al.}(2016)\citenamefont {Kar},
  \citenamefont {Hamad}, \citenamefont {Prasad}, \citenamefont {Cerchez},
  \citenamefont {Brauckmann}, \citenamefont {Aurand}, \citenamefont {Cantono},
  \citenamefont {Hadjisolomou}, \citenamefont {Lewis}, \citenamefont {Macchi},
  \citenamefont {Nersisyan}, \citenamefont {Robinson}, \citenamefont {Schroer},
  \citenamefont {Swantusch}, \citenamefont {Zepf}, \citenamefont {Willi},\ and\
  \citenamefont {Borghesi}}]{Nature2016}%
  \BibitemOpen
  \bibfield  {author} {\bibinfo {author} {\bibfnamefont {S.}~\bibnamefont
  {Kar}}, \bibinfo {author} {\bibfnamefont {A.}~\bibnamefont {Hamad}}, \bibinfo
  {author} {\bibfnamefont {R.}~\bibnamefont {Prasad}}, \bibinfo {author}
  {\bibfnamefont {M.}~\bibnamefont {Cerchez}}, \bibinfo {author} {\bibfnamefont
  {S.}~\bibnamefont {Brauckmann}}, \bibinfo {author} {\bibfnamefont
  {B.}~\bibnamefont {Aurand}}, \bibinfo {author} {\bibfnamefont
  {G.}~\bibnamefont {Cantono}}, \bibinfo {author} {\bibfnamefont
  {P.}~\bibnamefont {Hadjisolomou}}, \bibinfo {author} {\bibfnamefont
  {C.~L.~S.}\ \bibnamefont {Lewis}}, \bibinfo {author} {\bibfnamefont
  {A.}~\bibnamefont {Macchi}}, \bibinfo {author} {\bibfnamefont
  {G.}~\bibnamefont {Nersisyan}}, \bibinfo {author} {\bibfnamefont {A.~P.~L.}\
  \bibnamefont {Robinson}}, \bibinfo {author} {\bibfnamefont {A.~M.}\
  \bibnamefont {Schroer}}, \bibinfo {author} {\bibfnamefont {M.}~\bibnamefont
  {Swantusch}}, \bibinfo {author} {\bibfnamefont {M.}~\bibnamefont {Zepf}},
  \bibinfo {author} {\bibfnamefont {O.}~\bibnamefont {Willi}},\ and\ \bibinfo
  {author} {\bibfnamefont {M.}~\bibnamefont {Borghesi}},\ }\href
  {https://doi.org/https://doi.org/10.1038/ncomms10792} {\bibfield  {journal}
  {\bibinfo  {journal} {Nature Communications}\ }\textbf {\bibinfo {volume}
  {7}},\ \bibinfo {pages} {10792} (\bibinfo {year} {2016})}\BibitemShut
  {NoStop}%
\bibitem [{\citenamefont {Kumar}\ \emph {et~al.}(2019)\citenamefont {Kumar},
  \citenamefont {Gopal},\ and\ \citenamefont {Gupta}}]{kumar2019}%
  \BibitemOpen
  \bibfield  {author} {\bibinfo {author} {\bibfnamefont {S.}~\bibnamefont
  {Kumar}}, \bibinfo {author} {\bibfnamefont {K.}~\bibnamefont {Gopal}},\ and\
  \bibinfo {author} {\bibfnamefont {D.~N.}\ \bibnamefont {Gupta}},\ }\href@noop
  {} {\bibfield  {journal} {\bibinfo  {journal} {Plasma Physics and Controlled
  Fusion}\ }\textbf {\bibinfo {volume} {61}},\ \bibinfo {pages} {085001}
  (\bibinfo {year} {2019})}\BibitemShut {NoStop}%
\bibitem [{\citenamefont {Wilks}\ \emph {et~al.}(2001)\citenamefont {Wilks},
  \citenamefont {Langdon}, \citenamefont {Cowan}, \citenamefont {Roth},
  \citenamefont {Singh}, \citenamefont {Hatchett}, \citenamefont {Key},
  \citenamefont {Pennington}, \citenamefont {MacKinnon},\ and\ \citenamefont
  {Snavely}}]{doi:10.1063/1.1333697}%
  \BibitemOpen
  \bibfield  {author} {\bibinfo {author} {\bibfnamefont {S.~C.}\ \bibnamefont
  {Wilks}}, \bibinfo {author} {\bibfnamefont {A.~B.}\ \bibnamefont {Langdon}},
  \bibinfo {author} {\bibfnamefont {T.~E.}\ \bibnamefont {Cowan}}, \bibinfo
  {author} {\bibfnamefont {M.}~\bibnamefont {Roth}}, \bibinfo {author}
  {\bibfnamefont {M.}~\bibnamefont {Singh}}, \bibinfo {author} {\bibfnamefont
  {S.}~\bibnamefont {Hatchett}}, \bibinfo {author} {\bibfnamefont {M.~H.}\
  \bibnamefont {Key}}, \bibinfo {author} {\bibfnamefont {D.}~\bibnamefont
  {Pennington}}, \bibinfo {author} {\bibfnamefont {A.}~\bibnamefont
  {MacKinnon}},\ and\ \bibinfo {author} {\bibfnamefont {R.~A.}\ \bibnamefont
  {Snavely}},\ }\href@noop {} {\bibfield  {journal} {\bibinfo  {journal}
  {Physics of Plasmas}\ }\textbf {\bibinfo {volume} {8}},\ \bibinfo {pages}
  {542} (\bibinfo {year} {2001})}\BibitemShut {NoStop}%
\bibitem [{\citenamefont {Snavely}\ \emph {et~al.}(2000)\citenamefont
  {Snavely}, \citenamefont {Key}, \citenamefont {Hatchett}, \citenamefont
  {Cowan}, \citenamefont {Roth}, \citenamefont {Phillips}, \citenamefont
  {Stoyer}, \citenamefont {Henry}, \citenamefont {Sangster}, \citenamefont
  {Singh}, \citenamefont {Wilks}, \citenamefont {MacKinnon}, \citenamefont
  {Offenberger}, \citenamefont {Pennington}, \citenamefont {Yasuike},
  \citenamefont {Langdon}, \citenamefont {Lasinski}, \citenamefont {Johnson},
  \citenamefont {Perry},\ and\ \citenamefont {Campbell}}]{PhysRevLett.85.2945}%
  \BibitemOpen
  \bibfield  {author} {\bibinfo {author} {\bibfnamefont {R.~A.}\ \bibnamefont
  {Snavely}}, \bibinfo {author} {\bibfnamefont {M.~H.}\ \bibnamefont {Key}},
  \bibinfo {author} {\bibfnamefont {S.~P.}\ \bibnamefont {Hatchett}}, \bibinfo
  {author} {\bibfnamefont {T.~E.}\ \bibnamefont {Cowan}}, \bibinfo {author}
  {\bibfnamefont {M.}~\bibnamefont {Roth}}, \bibinfo {author} {\bibfnamefont
  {T.~W.}\ \bibnamefont {Phillips}}, \bibinfo {author} {\bibfnamefont {M.~A.}\
  \bibnamefont {Stoyer}}, \bibinfo {author} {\bibfnamefont {E.~A.}\
  \bibnamefont {Henry}}, \bibinfo {author} {\bibfnamefont {T.~C.}\ \bibnamefont
  {Sangster}}, \bibinfo {author} {\bibfnamefont {M.~S.}\ \bibnamefont {Singh}},
  \bibinfo {author} {\bibfnamefont {S.~C.}\ \bibnamefont {Wilks}}, \bibinfo
  {author} {\bibfnamefont {A.}~\bibnamefont {MacKinnon}}, \bibinfo {author}
  {\bibfnamefont {A.}~\bibnamefont {Offenberger}}, \bibinfo {author}
  {\bibfnamefont {D.~M.}\ \bibnamefont {Pennington}}, \bibinfo {author}
  {\bibfnamefont {K.}~\bibnamefont {Yasuike}}, \bibinfo {author} {\bibfnamefont
  {A.~B.}\ \bibnamefont {Langdon}}, \bibinfo {author} {\bibfnamefont {B.~F.}\
  \bibnamefont {Lasinski}}, \bibinfo {author} {\bibfnamefont {J.}~\bibnamefont
  {Johnson}}, \bibinfo {author} {\bibfnamefont {M.~D.}\ \bibnamefont {Perry}},\
  and\ \bibinfo {author} {\bibfnamefont {E.~M.}\ \bibnamefont {Campbell}},\
  }\href@noop {} {\bibfield  {journal} {\bibinfo  {journal} {Physical Review
  Letters}\ }\textbf {\bibinfo {volume} {85}},\ \bibinfo {pages} {2945}
  (\bibinfo {year} {2000})}\BibitemShut {NoStop}%
\bibitem [{\citenamefont {Hegelich}\ \emph {et~al.}(2005)\citenamefont
  {Hegelich}, \citenamefont {Albright}, \citenamefont {Audebert}, \citenamefont
  {Blazevic}, \citenamefont {Brambrink}, \citenamefont {Cobble}, \citenamefont
  {Cowan}, \citenamefont {Fuchs}, \citenamefont {Gauthier}, \citenamefont
  {Gautier}, \citenamefont {Geissel}, \citenamefont {Habs}, \citenamefont
  {Johnson}, \citenamefont {Karsch}, \citenamefont {Kemp}, \citenamefont
  {Letzring}, \citenamefont {Roth}, \citenamefont {Schramm}, \citenamefont
  {Schreiber}, \citenamefont {Witte},\ and\ \citenamefont
  {Fern{\'a}ndez}}]{hegelich2005}%
  \BibitemOpen
  \bibfield  {author} {\bibinfo {author} {\bibfnamefont {B.~M.}\ \bibnamefont
  {Hegelich}}, \bibinfo {author} {\bibfnamefont {B.}~\bibnamefont {Albright}},
  \bibinfo {author} {\bibfnamefont {P.}~\bibnamefont {Audebert}}, \bibinfo
  {author} {\bibfnamefont {A.}~\bibnamefont {Blazevic}}, \bibinfo {author}
  {\bibfnamefont {E.}~\bibnamefont {Brambrink}}, \bibinfo {author}
  {\bibfnamefont {J.}~\bibnamefont {Cobble}}, \bibinfo {author} {\bibfnamefont
  {T.}~\bibnamefont {Cowan}}, \bibinfo {author} {\bibfnamefont
  {J.}~\bibnamefont {Fuchs}}, \bibinfo {author} {\bibfnamefont {J.~C.}\
  \bibnamefont {Gauthier}}, \bibinfo {author} {\bibfnamefont {C.}~\bibnamefont
  {Gautier}}, \bibinfo {author} {\bibfnamefont {M.}~\bibnamefont {Geissel}},
  \bibinfo {author} {\bibfnamefont {D.}~\bibnamefont {Habs}}, \bibinfo {author}
  {\bibfnamefont {R.}~\bibnamefont {Johnson}}, \bibinfo {author} {\bibfnamefont
  {S.}~\bibnamefont {Karsch}}, \bibinfo {author} {\bibfnamefont
  {A.}~\bibnamefont {Kemp}}, \bibinfo {author} {\bibfnamefont {S.}~\bibnamefont
  {Letzring}}, \bibinfo {author} {\bibfnamefont {M.}~\bibnamefont {Roth}},
  \bibinfo {author} {\bibfnamefont {U.}~\bibnamefont {Schramm}}, \bibinfo
  {author} {\bibfnamefont {J.}~\bibnamefont {Schreiber}}, \bibinfo {author}
  {\bibfnamefont {K.~J.}\ \bibnamefont {Witte}},\ and\ \bibinfo {author}
  {\bibfnamefont {J.~C.}\ \bibnamefont {Fern{\'a}ndez}},\ }\href@noop {}
  {\bibfield  {journal} {\bibinfo  {journal} {Physics of Plasmas}\ }\textbf
  {\bibinfo {volume} {12}},\ \bibinfo {pages} {056314} (\bibinfo {year}
  {2005})}\BibitemShut {NoStop}%
\bibitem [{\citenamefont {Robinson}\ \emph {et~al.}(2008)\citenamefont
  {Robinson}, \citenamefont {Zepf}, \citenamefont {Kar}, \citenamefont
  {Evans},\ and\ \citenamefont {Bellei}}]{Robinson_2008}%
  \BibitemOpen
  \bibfield  {author} {\bibinfo {author} {\bibfnamefont {A.~P.~L.}\
  \bibnamefont {Robinson}}, \bibinfo {author} {\bibfnamefont {M.}~\bibnamefont
  {Zepf}}, \bibinfo {author} {\bibfnamefont {S.}~\bibnamefont {Kar}}, \bibinfo
  {author} {\bibfnamefont {R.~G.}\ \bibnamefont {Evans}},\ and\ \bibinfo
  {author} {\bibfnamefont {C.}~\bibnamefont {Bellei}},\ }\href@noop {}
  {\bibfield  {journal} {\bibinfo  {journal} {New Journal of Physics}\ }\textbf
  {\bibinfo {volume} {10}},\ \bibinfo {pages} {013021} (\bibinfo {year}
  {2008})}\BibitemShut {NoStop}%
\bibitem [{\citenamefont {Klimo}\ \emph {et~al.}(2008)\citenamefont {Klimo},
  \citenamefont {Psikal}, \citenamefont {Limpouch},\ and\ \citenamefont
  {Tikhonchuk}}]{klimo2008}%
  \BibitemOpen
  \bibfield  {author} {\bibinfo {author} {\bibfnamefont {O.}~\bibnamefont
  {Klimo}}, \bibinfo {author} {\bibfnamefont {J.}~\bibnamefont {Psikal}},
  \bibinfo {author} {\bibfnamefont {J.}~\bibnamefont {Limpouch}},\ and\
  \bibinfo {author} {\bibfnamefont {V.}~\bibnamefont {Tikhonchuk}},\
  }\href@noop {} {\bibfield  {journal} {\bibinfo  {journal} {Physical Review
  Special Topics-Accelerators and Beams}\ }\textbf {\bibinfo {volume} {11}},\
  \bibinfo {pages} {031301} (\bibinfo {year} {2008})}\BibitemShut {NoStop}%
\bibitem [{\citenamefont {Qiao}\ \emph {et~al.}(2012)\citenamefont {Qiao},
  \citenamefont {Kar}, \citenamefont {Geissler}, \citenamefont {Gibbon},
  \citenamefont {Zepf},\ and\ \citenamefont {Borghesi}}]{kar2012}%
  \BibitemOpen
  \bibfield  {author} {\bibinfo {author} {\bibfnamefont {B.}~\bibnamefont
  {Qiao}}, \bibinfo {author} {\bibfnamefont {S.}~\bibnamefont {Kar}}, \bibinfo
  {author} {\bibfnamefont {M.}~\bibnamefont {Geissler}}, \bibinfo {author}
  {\bibfnamefont {P.}~\bibnamefont {Gibbon}}, \bibinfo {author} {\bibfnamefont
  {M.}~\bibnamefont {Zepf}},\ and\ \bibinfo {author} {\bibfnamefont
  {M.}~\bibnamefont {Borghesi}},\ }\href@noop {} {\bibfield  {journal}
  {\bibinfo  {journal} {Physical Review Letters}\ }\textbf {\bibinfo {volume}
  {108}},\ \bibinfo {pages} {115002} (\bibinfo {year} {2012})}\BibitemShut
  {NoStop}%
\bibitem [{\citenamefont {Silva}\ \emph {et~al.}(2004)\citenamefont {Silva},
  \citenamefont {Marti}, \citenamefont {Davies}, \citenamefont {Fonseca},
  \citenamefont {Ren}, \citenamefont {Tsung},\ and\ \citenamefont
  {Mori}}]{silva2004}%
  \BibitemOpen
  \bibfield  {author} {\bibinfo {author} {\bibfnamefont {L.~O.}\ \bibnamefont
  {Silva}}, \bibinfo {author} {\bibfnamefont {M.}~\bibnamefont {Marti}},
  \bibinfo {author} {\bibfnamefont {J.~R.}\ \bibnamefont {Davies}}, \bibinfo
  {author} {\bibfnamefont {R.~A.}\ \bibnamefont {Fonseca}}, \bibinfo {author}
  {\bibfnamefont {C.}~\bibnamefont {Ren}}, \bibinfo {author} {\bibfnamefont
  {F.~S.}\ \bibnamefont {Tsung}},\ and\ \bibinfo {author} {\bibfnamefont
  {W.~B.}\ \bibnamefont {Mori}},\ }\href@noop {} {\bibfield  {journal}
  {\bibinfo  {journal} {Physical Review Letters}\ }\textbf {\bibinfo {volume}
  {92}},\ \bibinfo {pages} {015002} (\bibinfo {year} {2004})}\BibitemShut
  {NoStop}%
\bibitem [{\citenamefont {d'Humi{\'e}res}\ \emph {et~al.}(2005)\citenamefont
  {d'Humi{\'e}res}, \citenamefont {Lefebvre}, \citenamefont {Gremillet},\ and\
  \citenamefont {Malka}}]{d2005proton}%
  \BibitemOpen
  \bibfield  {author} {\bibinfo {author} {\bibfnamefont {E.}~\bibnamefont
  {d'Humi{\'e}res}}, \bibinfo {author} {\bibfnamefont {E.}~\bibnamefont
  {Lefebvre}}, \bibinfo {author} {\bibfnamefont {L.}~\bibnamefont
  {Gremillet}},\ and\ \bibinfo {author} {\bibfnamefont {V.}~\bibnamefont
  {Malka}},\ }\href@noop {} {\bibfield  {journal} {\bibinfo  {journal} {Physics
  of plasmas}\ }\textbf {\bibinfo {volume} {12}},\ \bibinfo {pages} {062704}
  (\bibinfo {year} {2005})}\BibitemShut {NoStop}%
\bibitem [{\citenamefont {Denavit}(1992)}]{PhysRevLett.69.3052}%
  \BibitemOpen
  \bibfield  {author} {\bibinfo {author} {\bibfnamefont {J.}~\bibnamefont
  {Denavit}},\ }\href@noop {} {\bibfield  {journal} {\bibinfo  {journal}
  {Physical Review Letters}\ }\textbf {\bibinfo {volume} {69}},\ \bibinfo
  {pages} {3052} (\bibinfo {year} {1992})}\BibitemShut {NoStop}%
\bibitem [{\citenamefont {Chen}\ \emph {et~al.}(2007)\citenamefont {Chen},
  \citenamefont {Sheng}, \citenamefont {Dong}, \citenamefont {He},
  \citenamefont {Li}, \citenamefont {Bari},\ and\ \citenamefont
  {Zhang}}]{chen2007}%
  \BibitemOpen
  \bibfield  {author} {\bibinfo {author} {\bibfnamefont {M.}~\bibnamefont
  {Chen}}, \bibinfo {author} {\bibfnamefont {Z.-M.}\ \bibnamefont {Sheng}},
  \bibinfo {author} {\bibfnamefont {Q.-L.}\ \bibnamefont {Dong}}, \bibinfo
  {author} {\bibfnamefont {M.-Q.}\ \bibnamefont {He}}, \bibinfo {author}
  {\bibfnamefont {Y.-T.}\ \bibnamefont {Li}}, \bibinfo {author} {\bibfnamefont
  {M.~A.}\ \bibnamefont {Bari}},\ and\ \bibinfo {author} {\bibfnamefont
  {J.}~\bibnamefont {Zhang}},\ }\href@noop {} {\bibfield  {journal} {\bibinfo
  {journal} {Physics of plasmas}\ }\textbf {\bibinfo {volume} {14}},\ \bibinfo
  {pages} {053102} (\bibinfo {year} {2007})}\BibitemShut {NoStop}%
\bibitem [{\citenamefont {He}\ \emph {et~al.}(2007)\citenamefont {He},
  \citenamefont {Dong}, \citenamefont {Sheng}, \citenamefont {Weng},
  \citenamefont {Chen}, \citenamefont {Wu},\ and\ \citenamefont
  {Zhang}}]{he2007}%
  \BibitemOpen
  \bibfield  {author} {\bibinfo {author} {\bibfnamefont {M.-Q.}\ \bibnamefont
  {He}}, \bibinfo {author} {\bibfnamefont {Q.-L.}\ \bibnamefont {Dong}},
  \bibinfo {author} {\bibfnamefont {Z.-M.}\ \bibnamefont {Sheng}}, \bibinfo
  {author} {\bibfnamefont {S.-M.}\ \bibnamefont {Weng}}, \bibinfo {author}
  {\bibfnamefont {M.}~\bibnamefont {Chen}}, \bibinfo {author} {\bibfnamefont
  {H.-C.}\ \bibnamefont {Wu}},\ and\ \bibinfo {author} {\bibfnamefont
  {J.}~\bibnamefont {Zhang}},\ }\href@noop {} {\bibfield  {journal} {\bibinfo
  {journal} {Physical Review E}\ }\textbf {\bibinfo {volume} {76}},\ \bibinfo
  {pages} {035402} (\bibinfo {year} {2007})}\BibitemShut {NoStop}%
\bibitem [{\citenamefont {Stockem}\ \emph {et~al.}(2013)\citenamefont
  {Stockem}, \citenamefont {Boella}, \citenamefont {Fiuza},\ and\ \citenamefont
  {Silva}}]{stockem2013}%
  \BibitemOpen
  \bibfield  {author} {\bibinfo {author} {\bibfnamefont {A.}~\bibnamefont
  {Stockem}}, \bibinfo {author} {\bibfnamefont {E.}~\bibnamefont {Boella}},
  \bibinfo {author} {\bibfnamefont {F.}~\bibnamefont {Fiuza}},\ and\ \bibinfo
  {author} {\bibfnamefont {L.}~\bibnamefont {Silva}},\ }\href@noop {}
  {\bibfield  {journal} {\bibinfo  {journal} {Physical Review E}\ }\textbf
  {\bibinfo {volume} {87}},\ \bibinfo {pages} {043116} (\bibinfo {year}
  {2013})}\BibitemShut {NoStop}%
\bibitem [{\citenamefont {Liu}\ \emph {et~al.}(2016)\citenamefont {Liu},
  \citenamefont {Weng}, \citenamefont {Li}, \citenamefont {Yuan}, \citenamefont
  {Chen}, \citenamefont {Mulser}, \citenamefont {Sheng}, \citenamefont
  {Murakami}, \citenamefont {Yu}, \citenamefont {Zheng},\ and\ \citenamefont
  {Zhang}}]{liu2016}%
  \BibitemOpen
  \bibfield  {author} {\bibinfo {author} {\bibfnamefont {M.}~\bibnamefont
  {Liu}}, \bibinfo {author} {\bibfnamefont {S.~M.}\ \bibnamefont {Weng}},
  \bibinfo {author} {\bibfnamefont {Y.~T.}\ \bibnamefont {Li}}, \bibinfo
  {author} {\bibfnamefont {D.~W.}\ \bibnamefont {Yuan}}, \bibinfo {author}
  {\bibfnamefont {M.}~\bibnamefont {Chen}}, \bibinfo {author} {\bibfnamefont
  {P.}~\bibnamefont {Mulser}}, \bibinfo {author} {\bibfnamefont {Z.~M.}\
  \bibnamefont {Sheng}}, \bibinfo {author} {\bibfnamefont {M.}~\bibnamefont
  {Murakami}}, \bibinfo {author} {\bibfnamefont {L.~L.}\ \bibnamefont {Yu}},
  \bibinfo {author} {\bibfnamefont {X.~L.}\ \bibnamefont {Zheng}},\ and\
  \bibinfo {author} {\bibfnamefont {J.}~\bibnamefont {Zhang}},\ }\href@noop {}
  {\bibfield  {journal} {\bibinfo  {journal} {Physics of Plasmas}\ }\textbf
  {\bibinfo {volume} {23}},\ \bibinfo {pages} {113103} (\bibinfo {year}
  {2016})}\BibitemShut {NoStop}%
\bibitem [{\citenamefont {Bhagawati}\ \emph {et~al.}(2019)\citenamefont
  {Bhagawati}, \citenamefont {Kuri},\ and\ \citenamefont
  {Das}}]{bhagawati2019}%
  \BibitemOpen
  \bibfield  {author} {\bibinfo {author} {\bibfnamefont {A.}~\bibnamefont
  {Bhagawati}}, \bibinfo {author} {\bibfnamefont {D.~K.}\ \bibnamefont
  {Kuri}},\ and\ \bibinfo {author} {\bibfnamefont {N.}~\bibnamefont {Das}},\
  }\href@noop {} {\bibfield  {journal} {\bibinfo  {journal} {Physics of
  Plasmas}\ }\textbf {\bibinfo {volume} {26}},\ \bibinfo {pages} {093106}
  (\bibinfo {year} {2019})}\BibitemShut {NoStop}%
\bibitem [{\citenamefont {Wei}\ \emph {et~al.}(2004)\citenamefont {Wei},
  \citenamefont {Mangles}, \citenamefont {Najmudin}, \citenamefont {Walton},
  \citenamefont {Gopal}, \citenamefont {Tatarakis}, \citenamefont {Dangor},
  \citenamefont {Clark}, \citenamefont {Evans}, \citenamefont {Fritzler}, ,
  \citenamefont {Clarke}, \citenamefont {Hernandez-Gomez}, \citenamefont
  {Neely}, \citenamefont {Mori}, \citenamefont {Tzoufras},\ and\ \citenamefont
  {Krushelnick}}]{wei2004ion}%
  \BibitemOpen
  \bibfield  {author} {\bibinfo {author} {\bibfnamefont {M.}~\bibnamefont
  {Wei}}, \bibinfo {author} {\bibfnamefont {S.}~\bibnamefont {Mangles}},
  \bibinfo {author} {\bibfnamefont {Z.}~\bibnamefont {Najmudin}}, \bibinfo
  {author} {\bibfnamefont {B.}~\bibnamefont {Walton}}, \bibinfo {author}
  {\bibfnamefont {A.}~\bibnamefont {Gopal}}, \bibinfo {author} {\bibfnamefont
  {M.}~\bibnamefont {Tatarakis}}, \bibinfo {author} {\bibfnamefont
  {A.}~\bibnamefont {Dangor}}, \bibinfo {author} {\bibfnamefont
  {E.}~\bibnamefont {Clark}}, \bibinfo {author} {\bibfnamefont
  {R.}~\bibnamefont {Evans}}, \bibinfo {author} {\bibfnamefont
  {S.}~\bibnamefont {Fritzler}}, , \bibinfo {author} {\bibfnamefont {R.~J.}\
  \bibnamefont {Clarke}}, \bibinfo {author} {\bibfnamefont {C.}~\bibnamefont
  {Hernandez-Gomez}}, \bibinfo {author} {\bibfnamefont {D.}~\bibnamefont
  {Neely}}, \bibinfo {author} {\bibfnamefont {W.}~\bibnamefont {Mori}},
  \bibinfo {author} {\bibfnamefont {M.}~\bibnamefont {Tzoufras}},\ and\
  \bibinfo {author} {\bibfnamefont {K.}~\bibnamefont {Krushelnick}},\
  }\href@noop {} {\bibfield  {journal} {\bibinfo  {journal} {Physical Review
  Letters}\ }\textbf {\bibinfo {volume} {93}},\ \bibinfo {pages} {155003}
  (\bibinfo {year} {2004})}\BibitemShut {NoStop}%
\bibitem [{\citenamefont {Zhang}\ \emph {et~al.}(2015)\citenamefont {Zhang},
  \citenamefont {Shen}, \citenamefont {Wang}, \citenamefont {Xu}, \citenamefont
  {Liu}, \citenamefont {Liang}, \citenamefont {Leng}, \citenamefont {Li},
  \citenamefont {Yan}, \citenamefont {Chen},\ and\ \citenamefont
  {Xu}}]{zhang2015}%
  \BibitemOpen
  \bibfield  {author} {\bibinfo {author} {\bibfnamefont {H.}~\bibnamefont
  {Zhang}}, \bibinfo {author} {\bibfnamefont {B.}~\bibnamefont {Shen}},
  \bibinfo {author} {\bibfnamefont {W.}~\bibnamefont {Wang}}, \bibinfo {author}
  {\bibfnamefont {Y.}~\bibnamefont {Xu}}, \bibinfo {author} {\bibfnamefont
  {Y.}~\bibnamefont {Liu}}, \bibinfo {author} {\bibfnamefont {X.}~\bibnamefont
  {Liang}}, \bibinfo {author} {\bibfnamefont {Y.}~\bibnamefont {Leng}},
  \bibinfo {author} {\bibfnamefont {R.}~\bibnamefont {Li}}, \bibinfo {author}
  {\bibfnamefont {X.}~\bibnamefont {Yan}}, \bibinfo {author} {\bibfnamefont
  {J.}~\bibnamefont {Chen}},\ and\ \bibinfo {author} {\bibfnamefont
  {Z.}~\bibnamefont {Xu}},\ }\href@noop {} {\bibfield  {journal} {\bibinfo
  {journal} {Physics of Plasmas}\ }\textbf {\bibinfo {volume} {22}},\ \bibinfo
  {pages} {013113} (\bibinfo {year} {2015})}\BibitemShut {NoStop}%
\bibitem [{\citenamefont {Sorasio}\ \emph {et~al.}(2006)\citenamefont
  {Sorasio}, \citenamefont {Marti}, \citenamefont {Fonseca},\ and\
  \citenamefont {Silva}}]{sorasio2006}%
  \BibitemOpen
  \bibfield  {author} {\bibinfo {author} {\bibfnamefont {G.}~\bibnamefont
  {Sorasio}}, \bibinfo {author} {\bibfnamefont {M.}~\bibnamefont {Marti}},
  \bibinfo {author} {\bibfnamefont {R.}~\bibnamefont {Fonseca}},\ and\ \bibinfo
  {author} {\bibfnamefont {L.}~\bibnamefont {Silva}},\ }\href@noop {}
  {\bibfield  {journal} {\bibinfo  {journal} {Physical Review Letters}\
  }\textbf {\bibinfo {volume} {96}},\ \bibinfo {pages} {045005} (\bibinfo
  {year} {2006})}\BibitemShut {NoStop}%
\bibitem [{\citenamefont {Haberberger}\ \emph {et~al.}(2012)\citenamefont
  {Haberberger}, \citenamefont {Tochitsky}, \citenamefont {Fiuza},
  \citenamefont {Gong}, \citenamefont {Fonseca}, \citenamefont {Silva},
  \citenamefont {Mori},\ and\ \citenamefont {Joshi}}]{haberberger2012}%
  \BibitemOpen
  \bibfield  {author} {\bibinfo {author} {\bibfnamefont {D.}~\bibnamefont
  {Haberberger}}, \bibinfo {author} {\bibfnamefont {S.}~\bibnamefont
  {Tochitsky}}, \bibinfo {author} {\bibfnamefont {F.}~\bibnamefont {Fiuza}},
  \bibinfo {author} {\bibfnamefont {C.}~\bibnamefont {Gong}}, \bibinfo {author}
  {\bibfnamefont {R.~A.}\ \bibnamefont {Fonseca}}, \bibinfo {author}
  {\bibfnamefont {L.~O.}\ \bibnamefont {Silva}}, \bibinfo {author}
  {\bibfnamefont {W.~B.}\ \bibnamefont {Mori}},\ and\ \bibinfo {author}
  {\bibfnamefont {C.}~\bibnamefont {Joshi}},\ }\href@noop {} {\bibfield
  {journal} {\bibinfo  {journal} {Nature Physics}\ }\textbf {\bibinfo {volume}
  {8}},\ \bibinfo {pages} {95} (\bibinfo {year} {2012})}\BibitemShut {NoStop}%
\bibitem [{\citenamefont {Fi{\'u}za}\ \emph {et~al.}(2012)\citenamefont
  {Fi{\'u}za}, \citenamefont {Stockem}, \citenamefont {Boella}, \citenamefont
  {Fonseca}, \citenamefont {Silva}, \citenamefont {Haberberger}, \citenamefont
  {Tochitsky}, \citenamefont {Gong}, \citenamefont {Mori},\ and\ \citenamefont
  {Joshi}}]{fiuza2012}%
  \BibitemOpen
  \bibfield  {author} {\bibinfo {author} {\bibfnamefont {F.}~\bibnamefont
  {Fi{\'u}za}}, \bibinfo {author} {\bibfnamefont {A.}~\bibnamefont {Stockem}},
  \bibinfo {author} {\bibfnamefont {E.}~\bibnamefont {Boella}}, \bibinfo
  {author} {\bibfnamefont {R.}~\bibnamefont {Fonseca}}, \bibinfo {author}
  {\bibfnamefont {L.}~\bibnamefont {Silva}}, \bibinfo {author} {\bibfnamefont
  {D.}~\bibnamefont {Haberberger}}, \bibinfo {author} {\bibfnamefont
  {S.}~\bibnamefont {Tochitsky}}, \bibinfo {author} {\bibfnamefont
  {C.}~\bibnamefont {Gong}}, \bibinfo {author} {\bibfnamefont {W.~B.}\
  \bibnamefont {Mori}},\ and\ \bibinfo {author} {\bibfnamefont
  {C.}~\bibnamefont {Joshi}},\ }\href@noop {} {\bibfield  {journal} {\bibinfo
  {journal} {Physical Review Letters}\ }\textbf {\bibinfo {volume} {109}},\
  \bibinfo {pages} {215001} (\bibinfo {year} {2012})}\BibitemShut {NoStop}%
\bibitem [{\citenamefont {Fiuza}\ \emph {et~al.}(2013)\citenamefont {Fiuza},
  \citenamefont {Stockem}, \citenamefont {Boella}, \citenamefont {Fonseca},
  \citenamefont {Silva}, \citenamefont {Haberberger}, \citenamefont
  {Tochitsky}, \citenamefont {Mori},\ and\ \citenamefont {Joshi}}]{fiuza2013}%
  \BibitemOpen
  \bibfield  {author} {\bibinfo {author} {\bibfnamefont {F.}~\bibnamefont
  {Fiuza}}, \bibinfo {author} {\bibfnamefont {A.}~\bibnamefont {Stockem}},
  \bibinfo {author} {\bibfnamefont {E.}~\bibnamefont {Boella}}, \bibinfo
  {author} {\bibfnamefont {R.}~\bibnamefont {Fonseca}}, \bibinfo {author}
  {\bibfnamefont {L.}~\bibnamefont {Silva}}, \bibinfo {author} {\bibfnamefont
  {D.}~\bibnamefont {Haberberger}}, \bibinfo {author} {\bibfnamefont
  {S.}~\bibnamefont {Tochitsky}}, \bibinfo {author} {\bibfnamefont
  {W.}~\bibnamefont {Mori}},\ and\ \bibinfo {author} {\bibfnamefont
  {C.}~\bibnamefont {Joshi}},\ }\href@noop {} {\bibfield  {journal} {\bibinfo
  {journal} {Physics of Plasmas}\ }\textbf {\bibinfo {volume} {20}},\ \bibinfo
  {pages} {056304} (\bibinfo {year} {2013})}\BibitemShut {NoStop}%
\bibitem [{\citenamefont {Palmer}\ \emph {et~al.}(2011)\citenamefont {Palmer},
  \citenamefont {Dover}, \citenamefont {Pogorelsky}, \citenamefont {Babzien},
  \citenamefont {Dudnikova}, \citenamefont {Ispiriyan}, \citenamefont
  {Polyanskiy}, \citenamefont {Schreiber}, \citenamefont {Shkolnikov},
  \citenamefont {Yakimenko},\ and\ \citenamefont {Najmudin}}]{palmer2011}%
  \BibitemOpen
  \bibfield  {author} {\bibinfo {author} {\bibfnamefont {C.~A.~J.}\
  \bibnamefont {Palmer}}, \bibinfo {author} {\bibfnamefont {N.~P.}\
  \bibnamefont {Dover}}, \bibinfo {author} {\bibfnamefont {I.}~\bibnamefont
  {Pogorelsky}}, \bibinfo {author} {\bibfnamefont {M.}~\bibnamefont {Babzien}},
  \bibinfo {author} {\bibfnamefont {G.~I.}\ \bibnamefont {Dudnikova}}, \bibinfo
  {author} {\bibfnamefont {M.}~\bibnamefont {Ispiriyan}}, \bibinfo {author}
  {\bibfnamefont {M.~N.}\ \bibnamefont {Polyanskiy}}, \bibinfo {author}
  {\bibfnamefont {J.}~\bibnamefont {Schreiber}}, \bibinfo {author}
  {\bibfnamefont {P.}~\bibnamefont {Shkolnikov}}, \bibinfo {author}
  {\bibfnamefont {V.}~\bibnamefont {Yakimenko}},\ and\ \bibinfo {author}
  {\bibfnamefont {Z.}~\bibnamefont {Najmudin}},\ }\href@noop {} {\bibfield
  {journal} {\bibinfo  {journal} {Physical Review Letters}\ }\textbf {\bibinfo
  {volume} {106}},\ \bibinfo {pages} {014801} (\bibinfo {year}
  {2011})}\BibitemShut {NoStop}%
\bibitem [{\citenamefont {Zhang}\ \emph {et~al.}(2017)\citenamefont {Zhang},
  \citenamefont {Shen}, \citenamefont {Wang}, \citenamefont {Zhai},
  \citenamefont {Li}, \citenamefont {Lu}, \citenamefont {Li}, \citenamefont
  {Xu}, \citenamefont {Wang}, \citenamefont {Liang},\ and\ \citenamefont
  {Leng}}]{zhang2017}%
  \BibitemOpen
  \bibfield  {author} {\bibinfo {author} {\bibfnamefont {H.}~\bibnamefont
  {Zhang}}, \bibinfo {author} {\bibfnamefont {B.}~\bibnamefont {Shen}},
  \bibinfo {author} {\bibfnamefont {W.}~\bibnamefont {Wang}}, \bibinfo {author}
  {\bibfnamefont {S.}~\bibnamefont {Zhai}}, \bibinfo {author} {\bibfnamefont
  {S.}~\bibnamefont {Li}}, \bibinfo {author} {\bibfnamefont {X.}~\bibnamefont
  {Lu}}, \bibinfo {author} {\bibfnamefont {J.}~\bibnamefont {Li}}, \bibinfo
  {author} {\bibfnamefont {R.}~\bibnamefont {Xu}}, \bibinfo {author}
  {\bibfnamefont {X.}~\bibnamefont {Wang}}, \bibinfo {author} {\bibfnamefont
  {X.}~\bibnamefont {Liang}},\ and\ \bibinfo {author} {\bibfnamefont
  {Y.}~\bibnamefont {Leng}},\ }\href@noop {} {\bibfield  {journal} {\bibinfo
  {journal} {Physical Review Letters}\ }\textbf {\bibinfo {volume} {119}},\
  \bibinfo {pages} {164801} (\bibinfo {year} {2017})}\BibitemShut {NoStop}%
\bibitem [{\citenamefont {Arber}\ \emph {et~al.}(2015)\citenamefont {Arber},
  \citenamefont {Bennett}, \citenamefont {Brady}, \citenamefont
  {Lawrence-Douglas}, \citenamefont {Ramsay}, \citenamefont {Sircombe},
  \citenamefont {Gillies}, \citenamefont {Evans}, \citenamefont {Schmitz},
  \citenamefont {Bell},\ and\ \citenamefont {Ridgers}}]{arber2015}%
  \BibitemOpen
  \bibfield  {author} {\bibinfo {author} {\bibfnamefont {T.}~\bibnamefont
  {Arber}}, \bibinfo {author} {\bibfnamefont {K.}~\bibnamefont {Bennett}},
  \bibinfo {author} {\bibfnamefont {C.}~\bibnamefont {Brady}}, \bibinfo
  {author} {\bibfnamefont {A.}~\bibnamefont {Lawrence-Douglas}}, \bibinfo
  {author} {\bibfnamefont {M.}~\bibnamefont {Ramsay}}, \bibinfo {author}
  {\bibfnamefont {N.}~\bibnamefont {Sircombe}}, \bibinfo {author}
  {\bibfnamefont {P.}~\bibnamefont {Gillies}}, \bibinfo {author} {\bibfnamefont
  {R.}~\bibnamefont {Evans}}, \bibinfo {author} {\bibfnamefont
  {H.}~\bibnamefont {Schmitz}}, \bibinfo {author} {\bibfnamefont
  {A.}~\bibnamefont {Bell}},\ and\ \bibinfo {author} {\bibfnamefont
  {C.}~\bibnamefont {Ridgers}},\ }\href@noop {} {\bibfield  {journal} {\bibinfo
   {journal} {Plasma Physics and Controlled Fusion}\ }\textbf {\bibinfo
  {volume} {57}},\ \bibinfo {pages} {113001} (\bibinfo {year}
  {2015})}\BibitemShut {NoStop}%
\end{thebibliography}
\end{document}